\documentclass[iop]{emulateapj}
\usepackage{graphicx}
\usepackage{amsmath}
\usepackage{natbib}
\usepackage{pdfpages}

\begin{document}

\slugcomment{to be submitted to ApJ}
\shortauthors{Harris et al.}

\title{Globular Cluster Systems in Brightest Cluster Galaxies. III:  Beyond Bimodality}
        
\author{William E.~Harris\altaffilmark{1}, 
	 Stephanie M.~Ciccone\altaffilmark{1},
	 Gwendolyn M.~Eadie\altaffilmark{1},
        Oleg Y.~Gnedin\altaffilmark{2}, 
        Douglas Geisler\altaffilmark{3}, 
        Barry Rothberg\altaffilmark{4}, and
	Jeremy Bailin\altaffilmark{5}
%        John P.~Blakeslee\altaffilmark{2},
%        Bradley C.~Whitmore\altaffilmark{3}, 
}

\altaffiltext{1}{Department of Physics \& Astronomy, McMaster University, Hamilton, ON, Canada; 
harris@physics.mcmaster.ca, ciccons@mcmaster.ca, eadiegm@mcmaster.ca}
%\altaffiltext{2}{Herzberg Institute of Astrophysics, National Research Council of Canada, Victoria, BC V9E 2E7, Canada; jblakeslee@nrc-cnrc.gc.ca, patrick.cote@nrc-cnrc.gc.ca}
%\altaffiltext{3}{Space Telescope Science Institute, 3700 San Martin Drive, Baltimore MD 21218, USA; whitmore@stsci.edu}
\altaffiltext{2}{Department of Astronomy, University of Michigan, Ann Arbor, MI 48109; ognedin@umich.edu}
\altaffiltext{3}{Departamento de Astronomi\'a, Universidad de Concepci\'on, Casilla 160-C, Concepci\'on, Chile; dgeisler@astroudec.cl}
\altaffiltext{4}{LBT Observatory, University of Arizona, 933 N.Cherry Ave, Tucson AZ 85721, USA; dr.barry.rothberg@gmail.com}
\altaffiltext{5}{Department of Physics and Astronomy, University of Alabama, Box 870324, Tuscaloosa, AL 35487-0324, USA; jbailin@ua.edu}
%\altaffiltext{6}{Department of Astronomy, Peking University, Beijing 100871, China; peng@bac.pku.edu.cn}

\date{\today}

\begin{abstract}
	We present new deep photometry of the rich globular cluster (GC) systems around the
	Brightest Cluster Galaxies UGC 9799 (Abell 2052) and UGC 10143 (Abell 2147),
	obtained with the HST ACS and WFC3 cameras.  For comparison, we also present new
	reductions of similar HST/ACS data for the Coma supergiants NGC 4874 and 4889.
	All four of these galaxies have huge cluster populations 
	(to the radial limits of our data, comprising from 12000 to 23000 clusters per galaxy).  
	The metallicity distribution functions (MDFs)
	of the GCs can still be matched by a bimodal-Gaussian form where the metal-rich
	and metal-poor modes are separated by $\simeq 0.8$ dex, but the internal dispersions
	of each mode are so large that the total MDF becomes very broad and nearly continuous
	from [Fe/H] $\simeq -2.4$ to Solar.  There are, however, significant differences
	between galaxies in the relative numbers of \emph{metal-rich} clusters, suggesting
	that they underwent significantly different histories of mergers with
	massive, gas-rich halos.  Lastly, the proportion
	of metal-poor GCs rises especially rapidly outside projected radii $R \gtrsim 4 R_{eff}$,
	suggesting the importance of accreted dwarf satellites in the outer halo.
	Comprehensive models for the formation of GCs as part of the hierarchical formation
	of their parent galaxies will be needed to trace the systematic change in
	structure of the MDF with galaxy mass, from the distinctly bimodal form in
	smaller galaxies up to the broad continuum that we see in the very largest systems.
\end{abstract}


\section{Introduction}

As a preface to the themes of this paper, it is difficult to improve on the introduction
by \citet{geisler_etal1996}, which we quote directly:

\smallskip
\parbox{8.0cm}{\emph{``One of the major goals of modern astronomy is an understanding of galaxy formation.
An ideal tool for this study would be a witness which was both present at the long-since-vanished
first epoch when most galaxies formed, and yet still survives today to tell us its story.
In addition, we would like many such witnesses, to corroborate their stories, and we would like them to 
be easy to find.  Enter the globular clusters.  They are among our most powerful cosmological probes for 
investigating this key topic.''}} 

\smallskip

In this paper we continue an exploration of the globular cluster systems (GCSs) around Brightest 
Cluster Galaxies (BCGs), the central dominant objects in rich clusters of galaxies.
This BCG program extends our earlier work on central giant galaxies at distances within 100 Mpc
\citep{harris_etal2006,harris2009a} outward to richer galaxy-cluster environments and higher BCG luminosities.

Paper I of the current series \citep{harris_etal2014} presents new deep photometry obtained with the Hubble Space Telescope
(HST) cameras around seven BCGs at distances from 100 to 250 Mpc, 
along with an analysis of the luminosity functions (LFs) of their GCs.
We found the GCLFs to be strikingly similar in all systems, with trends that extend previous
analyses for smaller galaxies \citep{jordan2007,villegas2010} smoothly upward to the largest galaxy sizes known.
Paper II \citep{harris_etal2016} presents a more comprehensive analysis of the GCS around one of these
BCGs, NGC 6166 in Abell 2199, with the focus on its GCS metallicity distribution, spatial distribution, and total GC 
population. 

In this paper, we present similar data for two of the other BCGs in our program,
UGC 9799 (Abell 2052) and UGC 10143 (Abell 2147), as well as new data reductions for the two supergiant 
galaxies NGC 4874 and 4889 in the Coma cluster (Abell 1656), constructed from HST Archive data.
In all cases, the photometry reaches
very similar depths in absolute magnitude and employs the same color indices,
enabling homogeneous comparisons among all the systems.  As in Paper II, our focus in this paper is on
the metallicity and spatial distributions.  

In Section 2 a review of the 
literature for GC metallicity distributions is presented.  
In Section 3 and 4 the target galaxies and the photometric
reductions are described.  The color-magnitude diagrams for the GC systems are presented
in Section 5, and an analysis of the color (metallicity) distributions is presented in
Section 6 along with their notable galaxy-to-galaxy
differences.  The spatial distributions are discussed in Section 7.
In Sections 8 and 9 we discuss some implications of our findings and a brief summary.

Our complete photometric data for the 5 BCGs discussed here and in Paper II can be obtained from
the webpage address http://physwww.mcmaster.ca/~harris/BCGdata.html
or by request to the first author.

\section{Metallicity and Color Distributions}

A particularly informative feature of globular cluster (GC) populations in galaxies 
is their \emph{metallicity distribution function} (MDF).  In many 
galaxies, a long-standing empirical feature of the MDF is its \emph{bimodal} nature, with
a canonical metal-poor (MP) ``blue'' sequence centered 
near $\langle$[Fe/H]$\rangle \simeq -1.5$
and a metal-rich (MR) ``red'' sequence near $\langle$[Fe/H]$\rangle \simeq -0.5$. 
Many authors have adopted the view that this two-part structure of the MDF
is evidence for two major and perhaps distinct
star-forming epochs in the formation histories of large galaxies,
a view that has persisted for many years
\citep[e.g.][among many others]{zepf_ashman93,forbes_etal97,forbes_etal2011,brodie_strader06,arnold_etal2011,blom_etal2012a,cantiello_etal2014,brodie_etal2014,kartha_etal2016}.
However, bimodal MDFs are not characteristic of the field-halo stars in their parent
galaxies, in the few cases where it has been possible to compare the both GCs and halo stars
directly in the same galaxy 
\cite[e.g.][]{harris_etal2007,durrell_etal2010,rejkuba_etal2011,rejkuba_etal2014,monachesi_etal2016}. 
Reconciling this apparent mismatch between field stars and GCs presents an intriguing challenge for quantitative formation modelling.

As is the case for all GC work, the origins of the topic start with the Milky Way.
\citet{zinn85} clearly established the bimodal nature of the Milky Way GC 
population, finding that the cluster metallicities coupled closely with systematically
different kinematics and spatial distributions for the MP and MR subcomponents.  Gradually growing evidence for
these two Milky Way subsystems had accumulated in earlier papers
\citep[including among others][]{mayall1946,morgan1956,baade1958,kinman1959,marsakov_suchkov1976,searle_zinn1978,harris_canterna1979},
but culminated in Zinn's definitive analysis.

For distant galaxies, spectroscopically measured GC metallicities 
are observationally far more time-consuming to build up, and probing the full three-dimensional kinematics 
of the halo is out of reach.  Instead, GC integrated colors
are commonly used as proxies for metallicity, since large samples of GCs can be efficiently 
measured this way.  For very old and relatively simple stellar systems such as GCs,
integrated color is sensitive only to metallicity while
other factors such as mean age or CNO abundances have only second- or third-order effects.
The key empirical question is then how to convert the \emph{color distribution function}
(CDF) of a sample of GCs to its MDF and whether or not these different forms
are measuring the same thing.
The literature on this topic is extensive and continually developing, so
a full synthesis is probably still premature.  However, the issue seems to boil down
to two central and only partially related questions:
\begin{enumerate}
	\item  For GC systems, is the MDF intrinsically bimodal?
	\item  Does the CDF correctly represent the MDF shape after the appropriate transformation?
\end{enumerate}

The answer to the first question, based strictly on spectroscopic evidence,
now appears to be that bimodality is common but that there is no truly universal pattern.
As has been emphasized elsewhere \citep{strader_etal2011,usher_etal2012,brodie_etal2014},
spectroscopy of significant samples of GCs,
in many galaxies, is needed to go beyond the default assumption of bimodality confidently.
The Milky Way GC system is clearly bimodal (see the Appendix of Paper II for 
a recent version of its MDF based on high-dispersion spectroscopy measures of [Fe/H]).
But several other galaxies now have GC metallicity data constructed from spectrum line
strengths and for these, differing results emerge.

In M31, the nearest large galaxy, the GC MDF 
displays a broad and more uniformly populated distribution that is less easily matched
by a bimodal-Gaussian form
\citep{caldwell_romanowsky2016,caldwell_etal2011,cezario_etal2013,perrett_etal2002,barmby_etal2000},
which may reflect
the complex and extended growth history of the galaxy \citep{mcconnachie_etal09}.
Other large galaxies with well populated GC systems are rather well described
by bimodal, spectroscopic MDFs: these include NGC 5128 \citep{woodley_etal2010}, 
M81 \citep{ma_etal2013}, NGC 4472 \citep{strader_etal2007}, NGC 4594 \citep{alves-brito_etal2011},
NGC 3115 \citep{arnold_etal2011,brodie_etal2012}, and 8 other large normal ellipticals
\citep{foster_etal2010,usher_etal2012}.  However, notable 
galaxy-to-galaxy differences appear in the degree of overlap between the MP and 
MR `modes' and their internal dispersions.
Spectroscopically based evidence for trimodality or simply a more uniform [Fe/H] distribution
is indicated for M87 \citep{strader_etal2011}, NGC 4494 \citep{usher_etal2012}, and perhaps
NGC 4365 \citep{chies-santos_etal2012a,blom_etal2012b} as well as M31.

The answer to the second question -- how well the CDF represents the MDF -- 
depends strongly on which color index is being used.
Many indices from near-UV through to near-IR have now been tested and compared.  
Several discussions have claimed that a unimodal MDF is  
capable of being converted to a bimodal
CDF if the transformation is sufficiently nonlinear 
\citep[e.g.][]{yoon_etal2006,yoon_etal2011a,yoon_etal2011b,cantiello_blakeslee2007,chies-santos_etal2012b,kim_etal2013,chung_etal2016}.
As a numerical exercise this claim is certainly true, though
the continuing issue with these discussions is that they rely very heavily on 
single-stellar-population (SSP) theoretical modelling to develop translation
curves from [Fe/H] to a given color index. These model curves usually have quite complex
shapes, and the various available SSP libraries show notable disagreements
\citep[e.g.][]{peacock_etal2011,alves-brito_etal2011,brodie_etal2012,chung_etal2016}.  

By contrast, empirically based transformations from metallicity to color
that do not rely heavily on modelling tend to be much more nearly linear and thus
to yield CDFs that resemble the intrinsic MDFs rather well
\citep[e.g.][]{barmby_etal2000,caldwell_etal2016,usher_etal2012,usher_etal2015,peng_etal2006,spitler_etal2008,foster_etal2010,sinnott_etal2010,fan_etal2010,peacock_etal2011,vanderbeke_etal2014,brodie_etal2012,brodie_etal2014,cantiello_etal2014,sakari_wallerstein2016}.
SSP theoretical models, however, are useful for comparing the relative metallicity sensitivity and linearity of
different color indices.  The most effective ones include $(V-K), (B-I), (g'-i'), (g'-I), (g'-K)$, or $(C-T_1)$
\citep{barmby_etal2000,harris_etal2006,cantiello_blakeslee2007,spitler_etal2008,blakeslee_etal2012,forte_etal2013,kundu_zepf2007,sinnott_etal2010,fan_etal2010,foster_etal2010}.  
These indices combine high metallicity sensitivity with modest degrees of nonlinearity.
However, a large fraction of the CDFs available in the literature have been measured in $(V-I)$ 
(which is less sensitive to metallicity) or $(g'-z')$ (which is more nonlinear than others listed above)
\citep[e.g.][]{gebhardt_kissler-patig99,larsen_etal01,peng_etal2006,villegas2010}.
A diagnostic index that has gained more frequent use recently is the Ca triplet 
line strength, which correlates near-linearly with [Fe/H] in the
range [Fe/H] $\lesssim -0.5$ \citep{sakari_wallerstein2016,brodie_etal2012}.
At the highest metallicities, CaT may become less sensitive, but any change in slope
above [Fe/H] $\simeq -0.5$ will not be able to generate the intermediate-metallicity
``valley'' at [Fe/H] $\simeq -1$ between the normal MP and MR modes.

Interestingly, \citet{usher_etal2015} discuss evidence that different galaxies may have
different color-to-metallicity conversions.  \citet{cantiello_blakeslee2007} and 
\citet{forte_etal2013} note that for a given galaxy,
every color index should yield the same MDF if the
transformations are correct.  So far, applying this self-consistency test in practice
has rarely been possible.

A useful conclusion for the present seems to be that CDFs reflect the intrinsic shapes of
the MDFs, if the color indices being used are chosen well.  The extensive literature
that reveals clearly bimodal CDFs for many galaxies therefore continues to be important
\citep[e.g.][among many others]{geisler_etal1996,neilsen_tsvetanov1999,larsen_etal01,kundu_whitmore01,rhode_zepf2004,bassino2006,peng_etal2006,harris2009a,harris2009b,faifer_etal2011,jennings_etal2014,cantiello_etal2014,kartha_etal2014}.
At the same time, some galaxies are better described as trimodal, unimodal, or simply
broad without matching a simple Gaussian-type model
\citep[for specific examples, see][]{larsen_etal01,peng_etal2006,blom_etal2012a,larsen_etal2005,kundu_zepf2007,usher_etal2012}.
A valid theoretical model for GC formation in the larger context of galaxy formation must be
able to deal with this diversity of outcomes.

Starting with either the CDF or MDF,
the first empirical question is simply to establish how many 
components or ``modes'' are present
regardless of shape, and how similar these might be to the Milky Way;
or (alternately) whether or not a bimodal deconstruction is justified in the first place.
The first mention of a \emph{specifically Gaussian 
shape} for these modes that we are aware of is in \citet{zinn85}.  
\citet{zepf_ashman93} introduced a mixture-modelling numerical code (the 
since-popular KMM package) to make objective tests of unimodality (the null hypothesis) versus
multimodality, using the CDFs for the two giant ellipticals NGC 4472 and 5128 as testbed cases.
They concluded strongly that their CDFs are bimodal.  
A Gaussian model was implicitly used for fitting the components,
and, more or less by default, this quickly became the norm for later studies.  
Just a few years later, the ``bimodal Gaussian''
model was already rather firmly established in the literature of the subject
\citep[e.g.][]{geisler_etal1996}.  

\begin{figure}[t]
\vspace{-3.0cm}
\begin{center}
\includegraphics[width=0.5\textwidth]{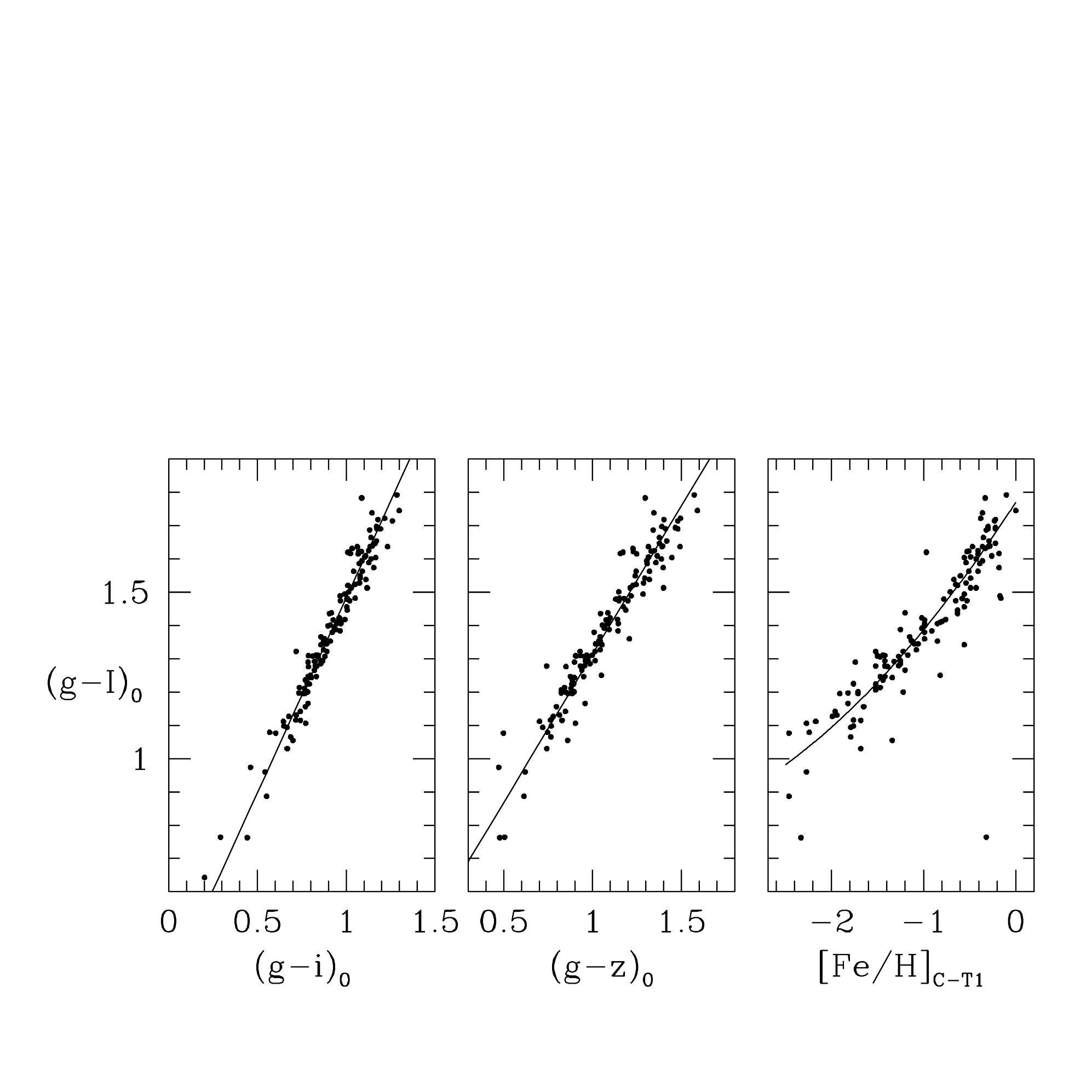}
\end{center}
\vspace{-0.2cm}
\caption{\emph{Left panel:} $(g-I)_0$ versus $(g-i)_0$ for clusters in NGC 5128.
	\emph{Middle panel:} $(g-I)_0$ versus $(g-Z)_0$ for the same clusters.
	\emph{Right panel:} $(g-I)_0$ versus [Fe/H] for the same clusters.
The equations for the interpolation lines are given in the text: in the right
panel, the solid line
is for [Fe/H] as the independent variable, and the dashed line for $(g-I)_0$
as the independent variable. Equation (1) in the text lies in between the two.}
\vspace{0.0cm}
\label{fig:gIcal}
\end{figure}

\citet{gebhardt_kissler-patig99} used various nonparametric tests to establish 
the Gaussian bimodal form more
strongly, though of necessity they
were forced to work with the rather metallicity-insensitive color index $(V-I)$ that 
dominated the available data at the time.  \citet{muratov_gnedin2010} discuss different
indicators of either Gaussianity or bimodality within the context of their GMM fitting
code, noting that a unimodal but asymmetric
MDF will often favor a bimodal-Gaussian fit.  
The bimodal-Gaussian form continues to be widely used simply because it continues
to match the CDFs and MDFs rather accurately
in large numbers of galaxies of all types, sizes, and environments
as the quality and internal precision of the data have steadily increased.
Nevertheless, there is not yet any \emph{a priori} astrophysical reason to say that
the MP and MR components should be specifically Gaussian. GC formation models within the
context of their parent galaxies are not yet advanced enough to make predictions for
the shape of the MDF at that level of precision \citep[e.g.][]{li_gnedin2014}.  
Thus the main purpose of these numerical model fits is to conveniently characterize
the first-order features of the MDF:  the mean metallicities of the modes (however many
there are), their widths (intrinsic metallicity spread), and the metallicity
separation between modes.

In the present survey of globular cluster systems in BCGs, we use the color index
$(F475W-F814W) \simeq (g-I)$ (from here on we drop the accents on the SDSS indices).  
In the following discussion it will be useful to have
a calibration of this index versus cluster metallicity [Fe/H].  To do this, we would
ideally need to
have GC photometry of the same clusters in both the Kron-Cousins and SDSS systems, in 
addition to spectroscopically based metallicity measurements.  At present, there are no
ideal solutions to that problem.  Galaxies satisfying all three
of these criteria are rare; in principle the Milky Way GC databases could be used, but
cluster-to-cluster foreground reddenings differ strongly, the published SDSS indices
\citep{vanderbeke_etal2014} show considerable scatter versus metallicity, 
and the variety of studies from which the $UBVRI$ indices
were derived are completely different from the SDSS survey, so aperture-size mismatches
are significant.  Similar problems affect the M31 GC sample.  The best 
option at the present time for developing a $(g-I)$ transformation is likely to be 
from the nearby early-type giant galaxy NGC 5128:
here, $UBVRI$ photometry is available from \citet{peng_etal2004}, $griz$ photometry 
from \citet{sinnott_etal2010}, and [Fe/H] values derived through $(C-T_1)$ from 
\citet{woodley_etal2010}; these [Fe/H] values are in turn well correlated with
the Sloan-system spectroscopic index [MgFe]' (see Woodley et al.).  We have extracted the GCs in common
from these three catalogs, with the results shown in Figure \ref{fig:gIcal}.  The great majority
of these GCs lie well outside the central few kiloparsecs of NGC 5128 and thus are 
unaffected by the well known dust lane.  We have therefore applied only the foreground
reddening of the galaxy, for which we adopt $E_{g-I} = 2.2 E_{B-V} = 0.25$ 
\citep{cardelli_etal1989} to obtain the intrinsic colors.  
We note, however, that the $UBVRI$ measurements were done on $3''$ aperture diameters corrected
to $14''$ through median curves of growth \citep{peng_etal2004}, while the $griz$ measures
were done through $7.6''$ apertures \citep{sinnott_etal2010}, so a small aperture mismatch
may exist here as well affecting the zeropoint of $(g-I)$.

The first two panels of Fig.~\ref{fig:gIcal} show the correlations between $(g-i)_0$ and
$(g-z)_0$ versus $(g-I)_0$.  These correlations rely purely on the 
photometric data independently of [Fe/H] estimates.  Simple linear relations
derived from direct least-squares fits are
\begin{eqnarray}
	(g-i)_0 \, & = & \, (-0.268 \pm 0.024) + (0.856 \pm 0.018) (g-I)_0 \nonumber \\
	(g-z)_0 \, & = & \, (-0.475 \pm 0.042) + (1.123 \pm 0.030) (g-I)_0 \, . \nonumber
\end{eqnarray}
The $(g-i)_0$ vs. $(g-I)_0$ relation is more tightly defined and valid over a wider range
of colors than $(g-z)_0$ vs. $(g-I)_0$.  
Both $(g-i)$ and $(g-I)$ do well as metallicity indicators, but $(g-I)$ is a bit more sensitive
and takes good advantage of the broadband HST filter system.
The $(g-z)$ index is in turn slightly more sensitive than
$(g-I)$, but a noticeable nonlinearity remains between them.
It should be noted again (see above) that the $g$ and $I$ photometric data come from two different observational
programs and therefore do not have the internal homogeneity that would normally be desired,
so any error in the zeropoint of the $(g-I)$ scale is hard to assess at present.  Fortunately, the slope and curvature 
of the relations are more important for the purposes here rather than the absolute values of [Fe/H].

The third panel of Fig.~\ref{fig:gIcal} connects $(g-I)_0$ with [Fe/H].
A modestly nonlinear quadratic relation accounting for scatter in both axes is
\begin{equation}
(g-I)_0 \, = \, 1.770 + 0.428 {\rm [Fe/H]} + 0.045 {\rm [Fe/H]}^2 \, 
\end{equation}
which is plotted in Fig.~\ref{fig:gIcal}c.  
We recognize that this proposed calibration is only temporary; in particular,
the zeropoint depends on the accuracy of the separate zeropoints of $g$ and $I$ from
two different observational programs and thus has a higher degree of uncertainty than usual.
The conversion of $(g-I)$ to [Fe/H] can be
greatly solidified once larger numbers of high-quality spectroscopically based 
[Fe/H] values become available for GC systems outside the Local Group particularly,
where aperture-size corrections on the photometry become unimportant.

\begin{table*}[t]
\begin{center}
\caption{\sc BCG Parameters}
\label{tab:basics}
\begin{tabular}{llcclcr}
\tableline\tableline\\
\multicolumn{1}{l}{Galaxy} &
\multicolumn{1}{l}{Cluster} &
\multicolumn{1}{c}{$(m-M)_I$} &
\multicolumn{1}{c}{$d$} &
\multicolumn{1}{l}{$M_K$} &
\multicolumn{1}{c}{$L_X$} &
\multicolumn{1}{r}{$R_c$ }
\\ & & & (Mpc) & & ($10^{44}$ erg s$^{-1}$) & (kpc) \\
\\[2mm] \tableline\\
NGC 4874 & A1656 & 35.02 & 100 & -26.1  & 3.98 & (0)  \\
NGC 4889 & A1656 & 35.02 & 100 & -25.6  & 3.98 & 169  \\
NGC 6166 & A2199 & 35.60 & 130 & -25.7  & 1.90 &   7  \\
UGC 9799 & A2052 & 35.95 & 150 & -25.5  & 1.33 &  38 \\
UGC 10143& A2147 & 35.99 & 154 & -24.9  & 1.66 &  82 \\
\\[2mm] \tableline
\end{tabular}
\end{center}
\vspace{0.4cm}
\end{table*}

\section{Target Galaxies}

In this paper, we present new photometry for the GC populations around UGC 9799 and UGC 10143. For 
comparison with these and the other BCGs in our program, we include as well NGC 4874 and NGC 4889
in the Coma cluster.  In all cases, the main data are from the ACS Wide Field Camera on board HST,
with identical filters and similar exposure time.  Here, we briefly summarize basic
features of these galaxies and the Abell clusters they dominate.

\begin{figure}[t]
\vspace{-0.0cm}
\begin{center}
\includegraphics[width=0.20\textwidth]{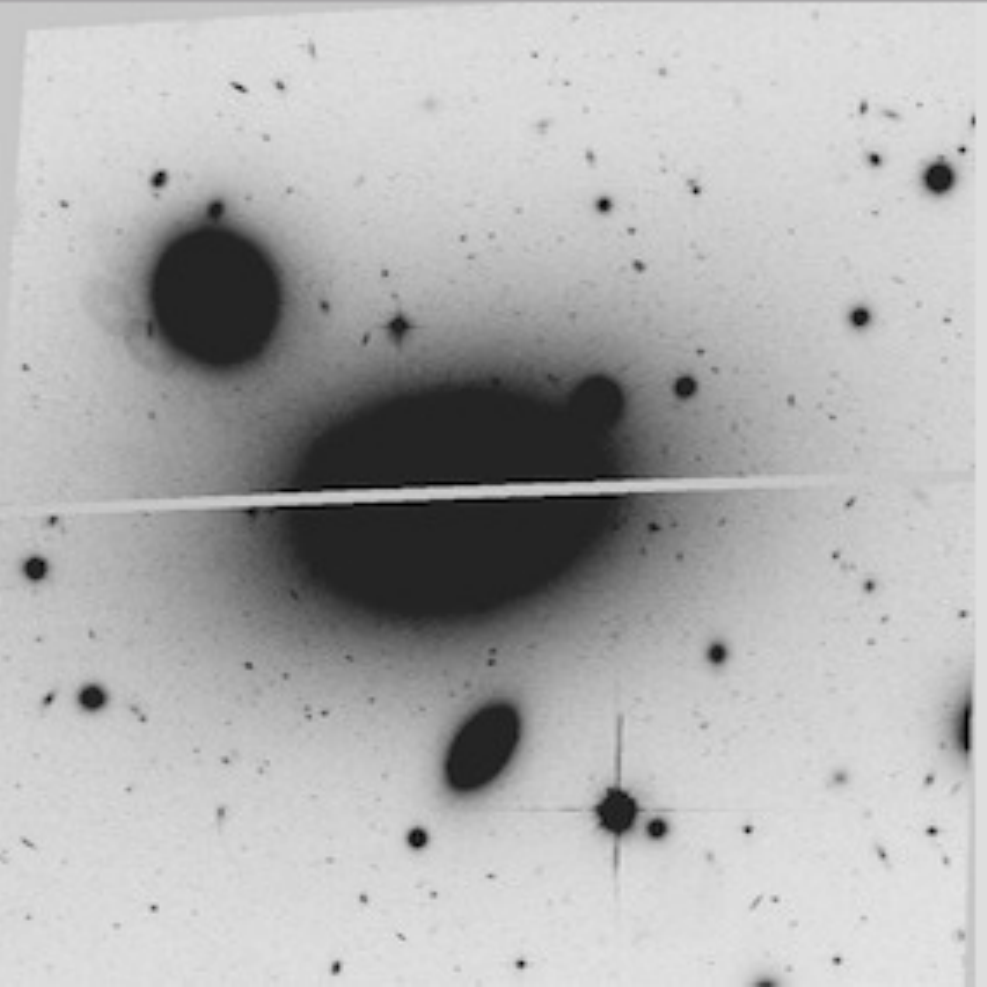}
\end{center}
\vspace{-0.2cm}
\caption{The ACS/WFC field centered on UGC 9799.  The field of
view is roughly 3.4 arcmin across.}
\vspace{0.0cm}
\label{fig:ugc9799}
\end{figure}

UGC 9799 is the central and brightest galaxy in Abell 2052 at a distance
$d = 150$ Mpc (for $H_0 = 70$ km s$^{-1}$ Mpc$^{-1}$).  As in Paper I we
adopt an apparent distance modulus $(m-M)_I = 35.95$ and foreground
reddening \citep[from NED, following][]{schlafly_finkbeiner2011}  $E_{B-V} = 0.037$.  
Detection of the GC system around UGC 9799 was first done by \citet{harris_etal1995}
through deep CFHT imaging.  A reproduction of our ACS/WFC field is shown in
Figure \ref{fig:ugc9799}.

\begin{figure}[t]
\vspace{0.0cm}
\begin{center}
\includegraphics[width=0.20\textwidth]{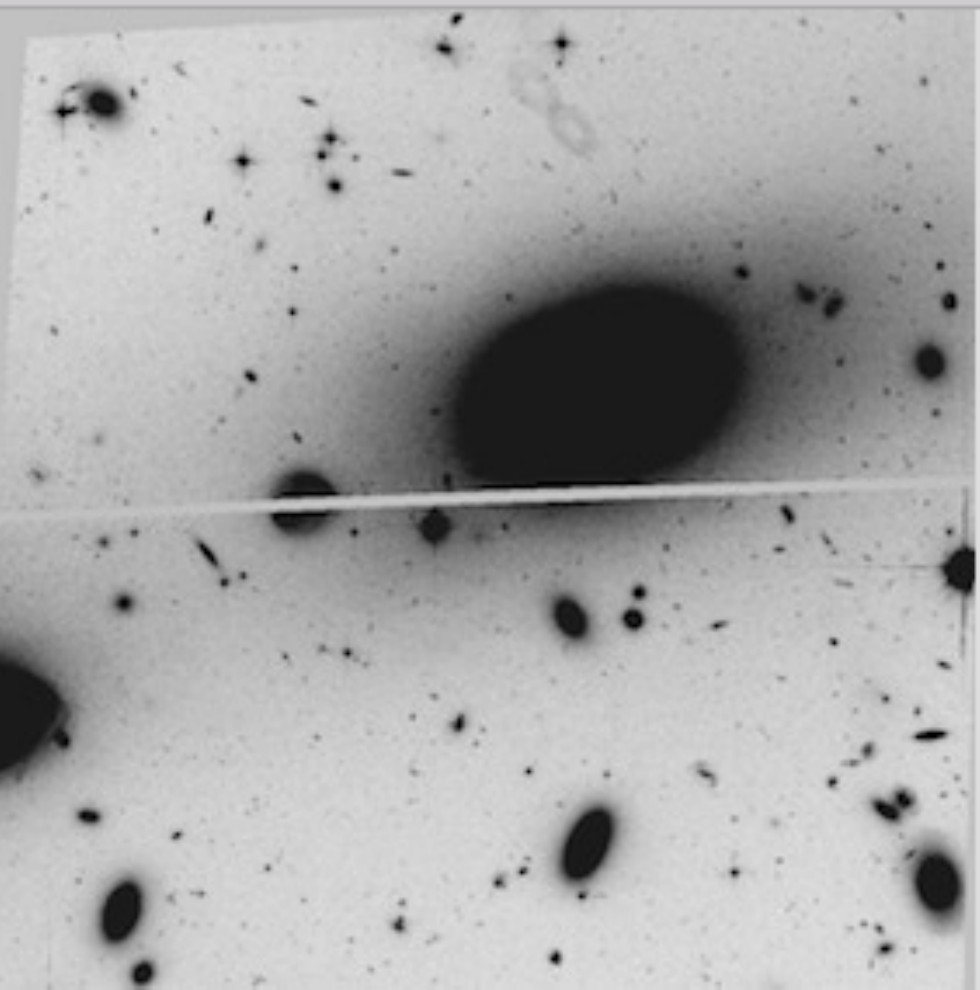}
\end{center}
\vspace{-0.2cm}
\caption{The ACS/WFC field centered on UGC 10143.} \vspace{0.0cm}
\label{fig:ugc10143}
\end{figure}

\begin{figure}[t]
\vspace{0.0cm}
\begin{center}
\includegraphics[width=0.20\textwidth]{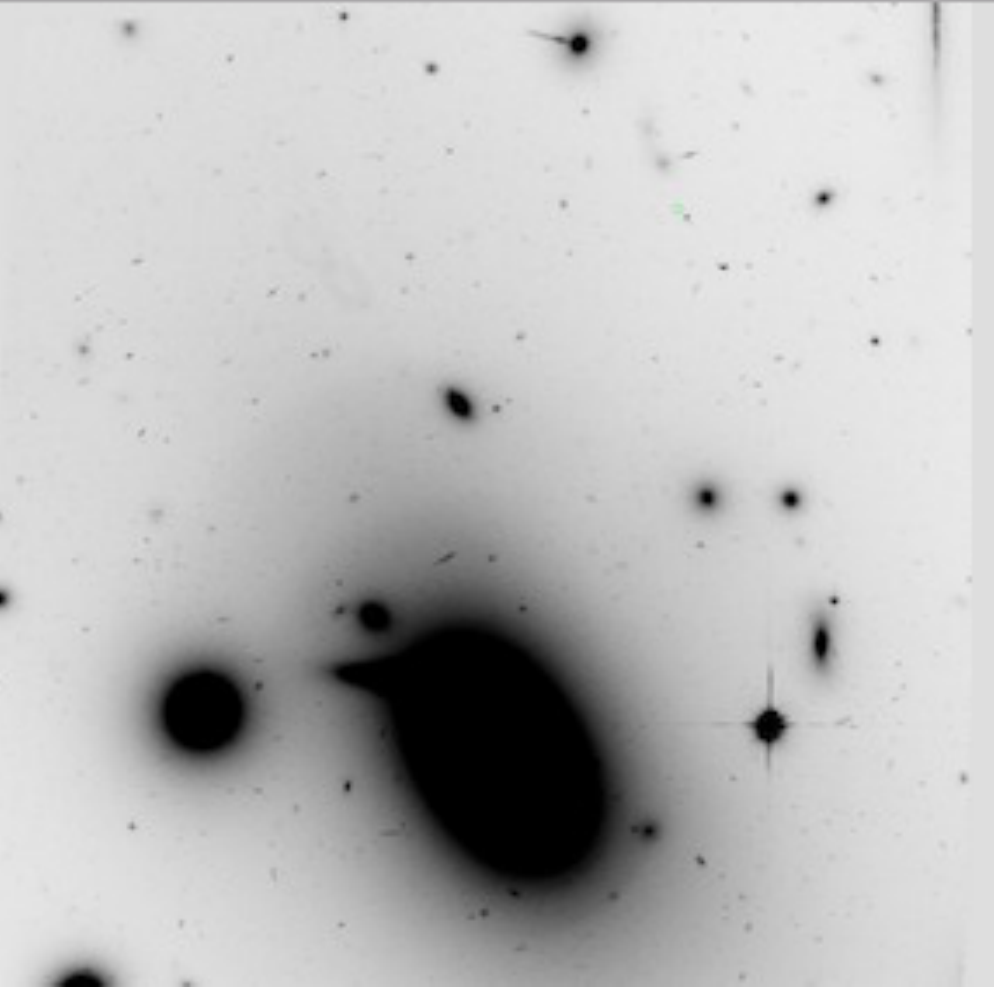}
\end{center}
\vspace{-0.2cm}
\caption{The ACS/WFC field containing NGC 4889.} \vspace{0.0cm}
\label{fig:ngc4889}
\end{figure}

\begin{figure}[t]
\vspace{0.0cm}
\begin{center}
\includegraphics[width=0.40\textwidth]{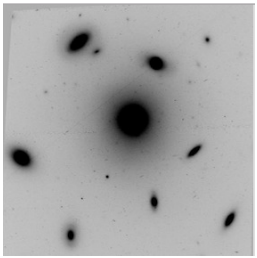}
\end{center}
\vspace{-0.2cm}
\caption{The ACS/WFC field (F2) centered on NGC 4874.} \vspace{0.0cm}
\label{fig:ngc4874}
\end{figure}

The core of UGC 9799 shows clear evidence for gas and (modest) nuclear activity.
As a radio source it is 3C 317, and Chandra observations reveal a compact core
X-ray source \citep{donato_etal2004,balmaverde_etal2006}.  At the center, GALEX
ultraviolet profiles and infrared excess lead to a deduced present-day star formation
rate of $\sim 1 M_{\odot}$ y$^{-1}$  \citep{hicks_etal2010,hoffer_etal2012}, while 
nuclear optical emission lines and a small star-forming filament visible in the
near-UV further confirm star-forming activity \citep{buttiglione_etal2010,martel_etal2002};
the filament is also visible in our $F475W$ image.
There is presumably a central supermassive black hole (SMBH), but only an upper limit
of $4.6 \times 10^9 M_{\odot}$ exists for it \citep{dallabonta_etal2009}.
The larger-scale distribution of hot X-ray gas in A2052 is discussed recently by
\citet{blanton_etal2011} and \citet{machado_limaneto2015}; the hot gas shows much substructure
(bubbles, shocks) indicative of previous AGN activity and/or merging of galaxies.

UGC 10143 is the central giant in A2147, at $d = 154$ Mpc; and A2147 is part of the
Hercules Supercluster along with A2151 and A2152 \citep{barmby_huchra1998}.  We adopt 
$(m-M)_I = 35.99$ and $E_{B-V} = 0.031$.  The BCG is radio-quiet and has a deduced
central star formation rate consistent with zero from its lack of UV or infrared signature
\citep{hoffer_etal2012}.  Close inspection of our images shows a smooth, featureless optical
core with no indications of peculiarities.
Detection of the GC system around UGC 10143 was first done by
\citet{blakeslee1999} through ground-based imaging and surface brightness fluctuation (SBF) analysis.
A reproduction of our ACS/WFC field is shown in Figure \ref{fig:ugc10143}.

According to \citet{tovmassian_andernach2012}
A2147 is not a Bautz-Morgan class I cluster since the luminosity difference in $M_K$
between its first- and second-ranked galaxies is only 0.2 mag (the second-ranked member is
PGC056770, which lies 180 kpc to the south of UGC 10143).  However, the overall 
cluster richness, the velocity dispersion, and the moderately low peculiar motion
of the BCG are all typical of BM I clusters, so the classification remains a bit
ambiguous.  

\begin{table*}[t]
\begin{center}
\caption{\sc Exposure Times and Completeness Parameters}
\label{tab:t}
\begin{tabular}{llllrll}
\tableline\tableline\\
\multicolumn{1}{l}{Galaxy} &
\multicolumn{1}{l}{GO Program} &
\multicolumn{1}{l}{Detector} &
\multicolumn{1}{l}{Filter} &
\multicolumn{1}{r}{$t$(sec)} &
\multicolumn{1}{l}{$m_0$} &
\multicolumn{1}{l}{$\alpha$} 
\\[2mm] \tableline\\
UGC 9799 & 12238 & ACS/WFC & F475W & 7977 & 29.60 & 3.5 \\
		       &  &       & F814W & 5253 & 28.13 & 3.5 \\
  & & WFC3    & F475W & 8041 & 29.30 & 3.0 \\
	 &  &       & F814W & 5343 & 27.85 & 2.7 \\
UGC 10143& 12238 & ACS/WFC & F475W & 10726 & 29.45 & 3.5 \\
		&	&         & F814W & 5262 & 27.95 & 3.3 \\
  & & WFC3    & F475W & 10856 & 29.20 & 3.0 \\
	 &   &      & F814W & 5352 & 27.30 & 3.3 \\
NGC 4889 & 11711 & ACS/WFC & F475W & 4770 & 29.20 & 2.6  \\
		       & &         & F814W & 9960 & 28.00 & 3.2 \\
NGC 4874-F1 & 10861 & ACS/WFC & F475W & 2677 & 28.30 & 2.6 \\
			  & &        & F814W & 1400 & 27.10 & 3.2 \\
NGC 4874-F2 & 11711 & ACS/WFC & F475W & 2394 & 28.60 & 2.6 \\
				 & &        & F814W & 10425 & 28.00 & 3.2 \\
NGC 4874-F3 & 12918 & ACS/WFC & F475W & 2568 & 28.30 & 2.6 \\
			  & &        & F814W & 1400 & 26.90 & 3.2 \\
\\[2mm] \tableline
\end{tabular}
\end{center}
\vspace{0.4cm}
\end{table*}

The giants NGC 4874 and NGC 4889 are the dominant galaxies in the rich and well known Coma cluster
(A1656), for which we adopt $d = 100$ Mpc, $(m-M)_I = 35.02$, and 
$E_{B-V} = 0.01$.  
Reproductions of the NGC 4889 and NGC4874-F2 fields from the ACS/WFC camera are shown
in Figures \ref{fig:ngc4889} and \ref{fig:ngc4874}.
NGC 4874 is surrounded by a handful of smaller satellite galaxies 
\citep[see Fig.~1 of][for their identification numbers]{cho_etal2016}, though as will be seen
below these do not contribute noticeably to the overall GC population, with the exception
of a small excess around NGC 4873 (at left center in Fig.~\ref{fig:ngc4874}).
Though NGC 4874 and 4889 have similar $V-$band luminosities, NGC 4874 is clearly the one lying
at or near the center of the Coma potential well (as defined by the intracluster X-ray gas)
and has a cD-type envelope.  By contrast NGC 4889 resembles a structurally more normal elliptical though
with a supergiant-level luminosity.  On our images, both have smooth featureless isophotes all the way inward
to the galaxy center.  The first detections of GC populations around these
galaxies were done 30 years ago by \citet{harris1987} and \citet{thompson_valdes1987} 
through deep imaging with the CFHT.
Later ground-based imaging by \citet{blakeslee_tonry1995} and \citet{marin-franch_aparicio2002}
including SBF techniques verified that both galaxies had rich GC systems.   \citet{harris_etal09} 
presented homogeneous photometry of the GC systems in 5 Coma ellipticals including the two
supergiants, all with data from the HST WFPC2 camera.  These reached deep enough to gauge the GC luminosity
function turnover point and to obtain useful values for the GC specific frequencies in the
galaxies, but color indices ($V-I$ in this case) were not precise enough to clearly resolve the CDF and determine whether
or not these systems fall within the conventional bimodal pattern.

More recently, \citet{peng_etal2011} discussed the distribution of GCs throughout the Coma cluster,
using the HST/ACS Coma Cluster Treasury Survey imaging.
Their analysis shows that NGC 4874 is
essentially at the center of the GC distribution in Coma and that its own GC spatial profile makes a clear transition to
a newly discovered Intra-Galactic cluster (IGC) population, which becomes dominant beyond
a projected radius of $\sim 300$ kpc.
In the CDF, two modes (MP and MR) are clearly present and the MP mode is much more dominant for
the IGC.  Color-magnitude diagrams for
the GC populations around both NGC 4874 and 4889 from HST ACS imaging are presented
and discussed briefly 
by \citet{lee_jang2016a}, while \citet{cho_etal2016} complete a more comprehensive discussion
specifically for the NGC 4874 system now including 3-color $(g,I,H)$ photometry.  

A structural feature held in common by all four of the galaxy clusters discussed here
(Coma, A2052, A2147, and A2199 from Paper II) is a prominent X-ray halo gas component.
Of 60 nearby clusters listed by \citet{edwards_etal2007} selected
from the NOAO Fundamental Plane Survey, these four rank among the highest in X-ray
luminosity, though not all of them have strong cooling flows or central optical emission.
IGC populations have clearly been established to date only for Virgo and Coma
\citep{durrell_etal2014,peng_etal2011},  and more tentatively in Abell 1689
\citep{alamo-martinez_etal2013}, Abell 1185 \citep{west_etal2011}, 
and Abell 2744 \citep{lee_jang2016b}, and it is not yet known how well their presence correlates with
hot halo gas.  However, the X-ray halos demonstrate that the BCGs studied here all reside in
very massive potential wells defined by their surrounding clusters.  
The virial masses of these clusters as confirmed through galaxy velocity dispersions,
X-ray gas temperature, or weak lensing are typically $M_{200} \sim 10^{14} - 10^{15} M_{\odot}$
\citep[e.g.][]{falco_etal2014,wojtak_lokas2010,wen_etal2010,kubo_etal2007,lokas_etal2006,blanton_etal2003}.

In Table \ref{tab:basics}, we summarize
some of the fiducial properties of the BCGs discussed in this paper.
The last column gives the projected distance $R_c$ of each BCG from
the center of its Abell cluster (see the references cited above).

\section{Photometric Reductions}

Imaging for our program was done
with the $F475W$ and $F814W$ filters.  The resulting color index in the
native filters, $(F475W-F814W)$, is close to standard $(g-I)$ and is both 
metallicity-sensitive and nearly linearly correlated with metallicity (see Section 2).
The magnitude scale we adopt here, as in previous work \citep[][and Papers I and II]{harris2009a},
is on the VEGAMAG system.

The raw imaging data for UGC 9799 and 10143 are from HST program GO-12238 (PI Harris).
Design parameters for this program are summarized in Paper I; full details
of the photometric data reductions are laid out in Paper II, and we follow the
same procedures here.  For these two galaxies, ACS/WFC exposures were taken roughly
centered on the BCG, while Parallel exposures with WFC3/UVIS (in the same filters)
were taken simultaneously to give an offset field located in the outskirts of the 
galaxy cluster.  

From the \emph{*.flc} raw image files provided in the $HST$ Archive we constructed
a single combined image in each filter with \emph{stsdas/multidrizzle}.  We used
SourceExtractor \citep{bertin_arnouts1996} to detect candidate objects in
each field, and to do a preliminary rejection of nonstellar objects.  
From there we used the normal sequence of steps in \emph{iraf/daophot/allstar}
\citep{stetson1987} to complete the photometry from PSF (point-spread function) fitting
and to do further rejection of nonstellar objects from the goodness-of-fit $\chi$
parameter and the internal measurement uncertainties in each filter.
In all cases, the candidate GCs we are searching for are expected to be starlike in
structure for galaxies more distant than $d \gtrsim 80$ Mpc 
\citep[see Papers I and II as well as][]{harris2009a}.  This is an important 
advantage for our purposes, because it facilitates the removal
of the vast majority of the field contamination, which 
is dominated by faint, very small but resolved background galaxies.

Lastly, artificial-star tests were run with \emph{daophot/addstar} to quantify the detection
completeness fraction as a function of magnitude, $f(m)$, separately for each
target field and filter.  The $f(m)$ data were fit to a smooth curve of the form
\begin{equation}
	f(m) \, = \, {1 \over {1 + e^{\alpha (m - m_0)} }}
\end{equation}
as defined in Paper II.
Here $m_0$ represents the 50\% completeness level and $\alpha$ the steepness
of falloff as the curve passes through $m_0$.  Nominally, $f$
is also a function of the background light intensity and therefore the
projected galactocentric distance.  However, for these distant and rather
diffuse BCGs, at
$R \gtrsim 15''$ the surface brightness has already fallen to a low enough level
that the radial dependence beyond that radius is negligible (see Paper I for discussion).
In our following analysis, we do not use any of the raw data within $15''$
of the galaxy centers.
More detailed descriptions of the procedures and examples can be found
in Paper II and \citet{harris2009a}.
In Table \ref{tab:t}, we list in successive columns the galaxy name, GO program
ID from the HST Archive, camera, filter name, total exposure time, and completeness
function parameters.

\section{Color-Magnitude Diagrams}

\begin{figure*}[t]
\vspace{-5.0cm}
\begin{center}
\includegraphics[width=0.9\textwidth]{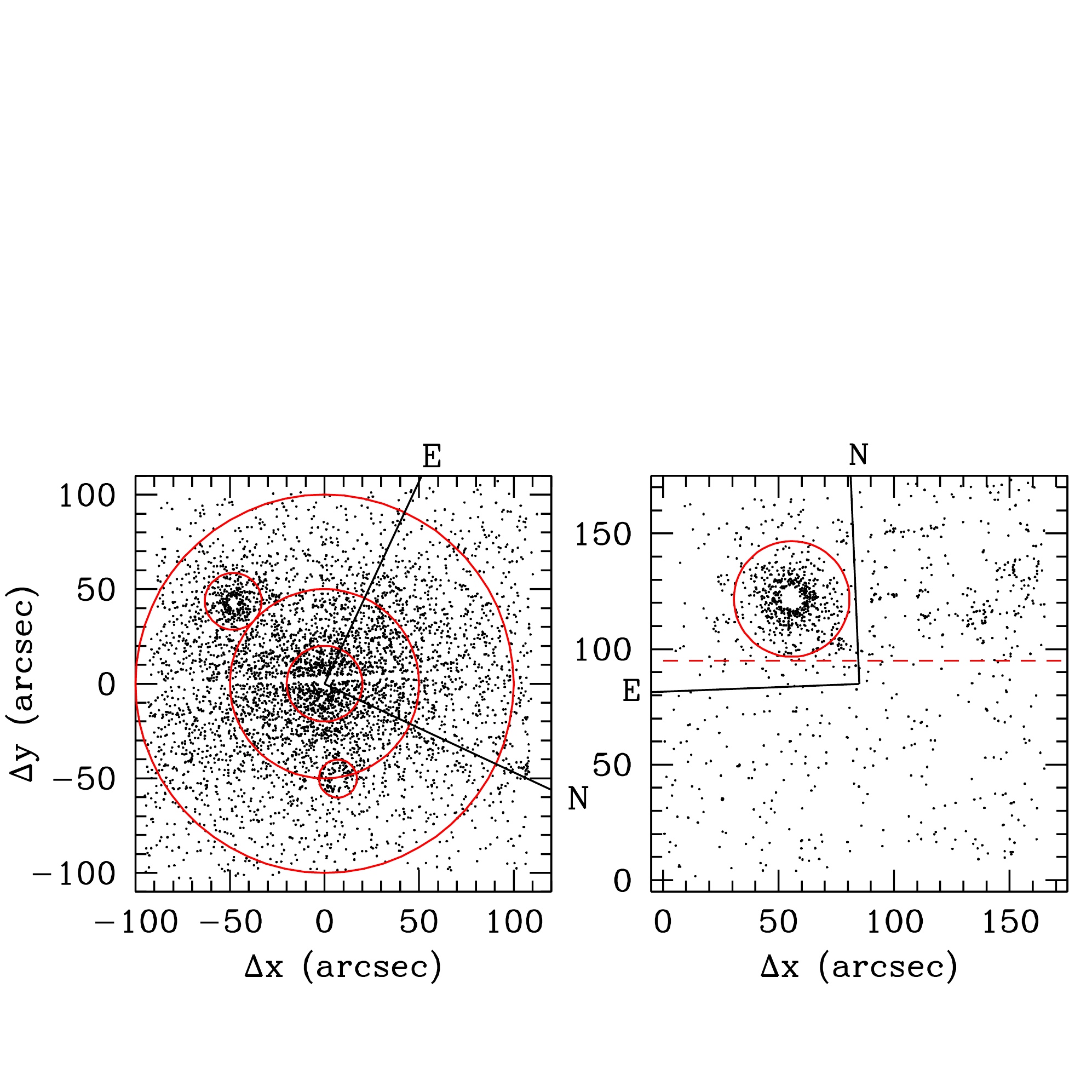}
\end{center}
\vspace{-1.5cm}
\caption{Locations of the measured starlike objects brighter
than $F814W = 27.0$ in the ACS field centered near
UGC 9799 (left panel) and in the Parallel WFC3 field (right panel).
Fiducial directions on the sky (North, East) are marked in both panels.
Small red circles mark smaller companion galaxies with GC populations
of their own, as listed in the text. For WFC3, the red dashed line indicates
the border between contaminating small galaxies (above the line) and a
cleaner sample of IGC GCs (below the line).}
\vspace{0.0cm}
\label{fig:ugc9799_xy}
\end{figure*}

In Figure \ref{fig:ugc9799_xy}, we show the distribution of \emph{measured starlike objects}
brighter than $F814W = 27.0$ 
in the UGC 9799 ACS and WFC3 fields.  In this magnitude range, as shown in Paper II
almost all of these objects are expected to
be GCs.  In the ACS field, two smaller companion galaxies are visible as separate
compact groups of GCs of their own:  these are PGC054528 (at upper left, marked out
by a circle of $15''$ radius), and PGC05421 (below UGC 9799 and marked out by a
$r=10''$ circle).  In the WFC3 field, the galaxy at upper left with an obvious GC
population of its own is PGC054533, marked by a $r=25''$ circle.  At upper right is a smaller elliptical
PGC054530.

\begin{figure*}[t]
\vspace{-5.0cm}
\begin{center}
\includegraphics[width=0.9\textwidth]{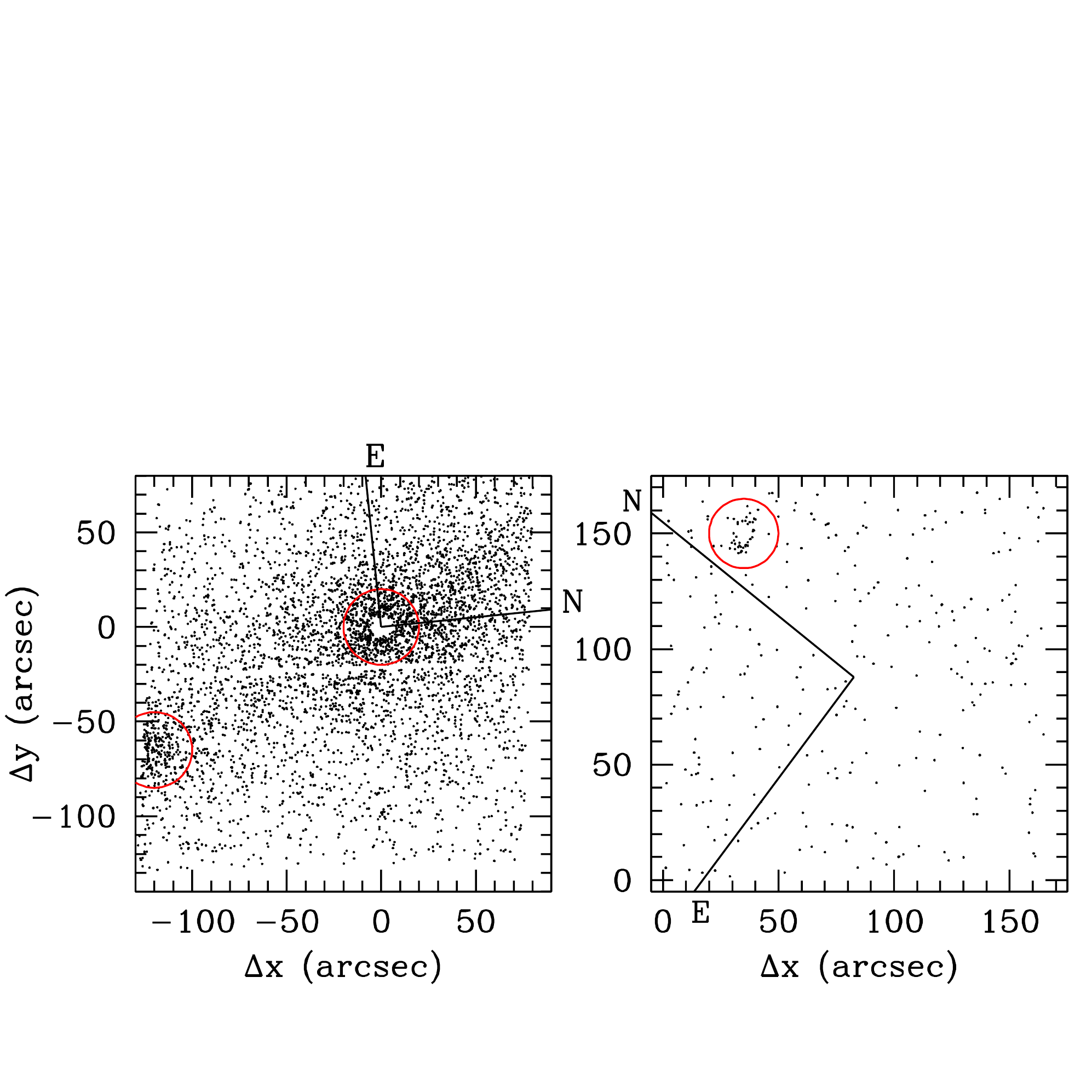}
\end{center}
\vspace{-1.5cm}
\caption{Locations of the measured starlike objects brighter
than $F814W = 27.0$ in the ACS field centered near
UGC 10143 (left panel) and in the Parallel WFC3 field (right panel).
In the ACS field the companion galaxy PGC056777 is shown by the $20''$
circle at lower left, while in the WFC3 field one small galaxy is marked
by a $15''$ circle at upper left (see text).}
\vspace{0.0cm}
\label{fig:ugc10143_xy}
\end{figure*}

In Figure \ref{fig:ugc10143_xy}, we show the distribution of measured starlike objects
with $F814W < 27.0$ in the UGC 10143 fields.
In the ACS field (left panel) one obvious clump of points at lower left marks PGC056777
($r=20''$ circle), a nearly face-on disk
galaxy with a complex and distorted array of spiral arms.
In the WFC3 field, only one relatively small galaxy appears at upper left ($r=15''$ circle), which is
2MASXJ16023373+1555259.

\begin{figure}[t]
\vspace{0.0cm}
\begin{center}
\includegraphics[width=0.5\textwidth]{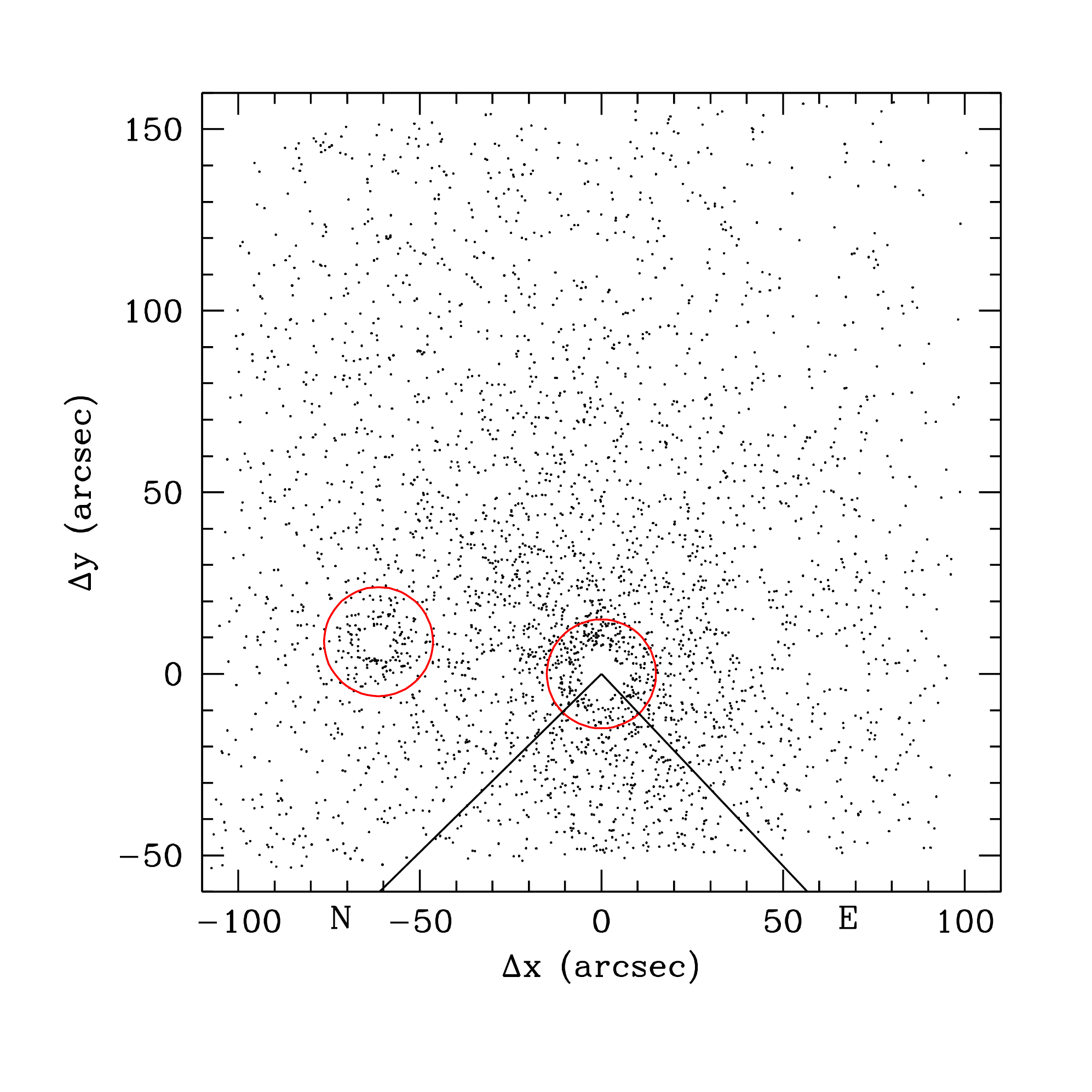}
\end{center}
\vspace{-0.5cm}
\caption{Locations of the measured starlike objects brighter
than $F814W = 26.5$ in the ACS field centered near
NGC 4889. A circle of $20''$ radius is marked around both the
center of NGC 4889 and the companion E0 galaxy NGC 4886 (at left).}
\vspace{0.0cm}
\label{fig:ngc4889_xy}
\end{figure}

\begin{figure}[t]
\vspace{0.0cm}
\begin{center}
\includegraphics[width=0.6\textwidth]{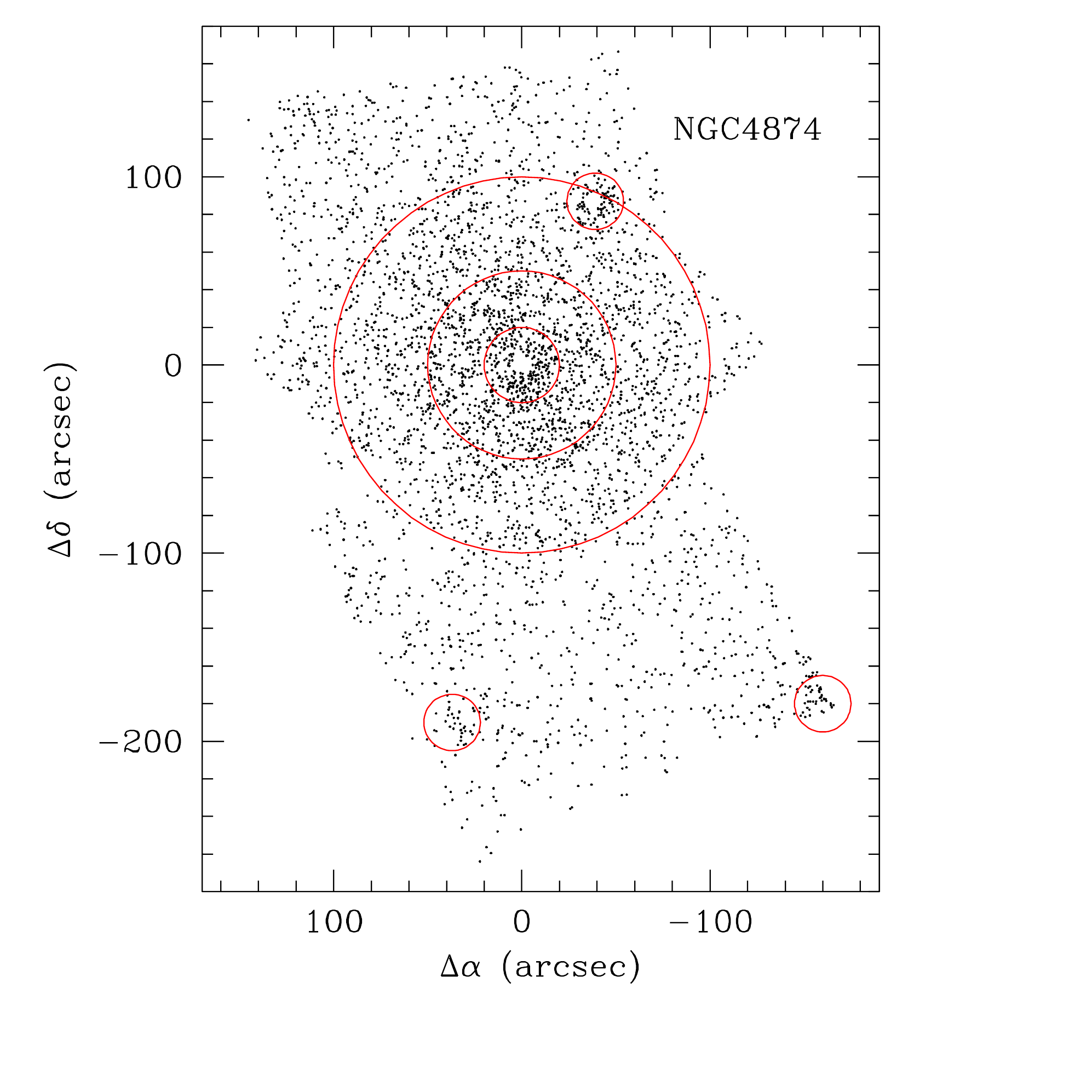}
\end{center}
\vspace{-1.0cm}
\caption{Locations of the measured starlike objects brighter
than $F814W = 26.0$ in the three overlapping ACS fields centered near
NGC 4874. Coordinates plotted are relative to the center of NGC 4874,
aligned following the cardinal axes 
with East at left and North at top.  Three small galaxies with noticeable GC populations of their
own are marked with red circles of $15''$ radius.  The concentric
circles around the center of NGC 4874 have radii of $20''$, $50''$, and
$100''$.}
\vspace{0.0cm}
\label{fig:ngc4874_xy}
\end{figure}

The same photometric procedures were followed for NGC 4889, with the resulting 
$xy$ plot as shown in Figure \ref{fig:ngc4889_xy}.  The clump of points to the 
left of the central giant galaxy indicates a GC population around the nearby
galaxy NGC 4886 (= NGC 4882), classified E0.  In between
these is the lenticular galaxy PGC044708, but this does not contribute significantly
to the GC population.

As noted above, the GC population around NGC 4874 has been analyzed by 
\citet{cho_etal2016} from the single field NGC4874-F2 as listed in Table
\ref{tab:t}.  To add a bit more statistical weight to our measurements
of this rich GC system and especially to increase the radial coverage, we added data
from two other overlapping fields with ACS exposures in the same filters 
(listed as NGC4874-F1 and NGC4874-F3 in Table \ref{tab:t}), though the exposures
in field F2 reach the deepest of the three (see below).\footnote{The
measurements by Cho et al.~are on the ABMAG system, whereas our independent reductions
are on the VEGAMAG system to make them strictly comparable with the other BCGs in our study.
Their photometry also uses SExtractor parameters for the photometry whereas our data are from \emph{daophot/allstar};
we used SE only for object detection and preliminary culling.}
To guarantee that all three fields
were on the same internal magnitude scale, we used the overlapping objects measured
in more than one field to define mean magnitude offsets in both filters and 
normalize fields F1 and F3 to the magnitude scales of F2.  These offsets were all less than
$\pm 0.03$ mag in either filter, which is within the internal uncertainties of the 
large-aperture corrections to the \emph{allstar} PSF-fitting magnitudes (see Paper II).
To define a final photometric dataset, we used F2, plus the regions of F1 and F3
that fall outside the area covered by F2.
The $xy$ plot for the three fields combined is shown in Figure \ref{fig:ngc4874_xy}.
As is evident in the Figure, the camera orientation angles were 
different for each field, leaving a somewhat irregularly shaped composite field.
The three smaller galaxies marked with $r=15''$ circles are NGC 4873 (at top),
NGC 4875 (lower left), and NGC 4869 (lower right), and these are excluded from
later analysis.

\begin{figure*}[t]
\vspace{-2.0cm}
\begin{center}
\includegraphics[width=0.6\textwidth]{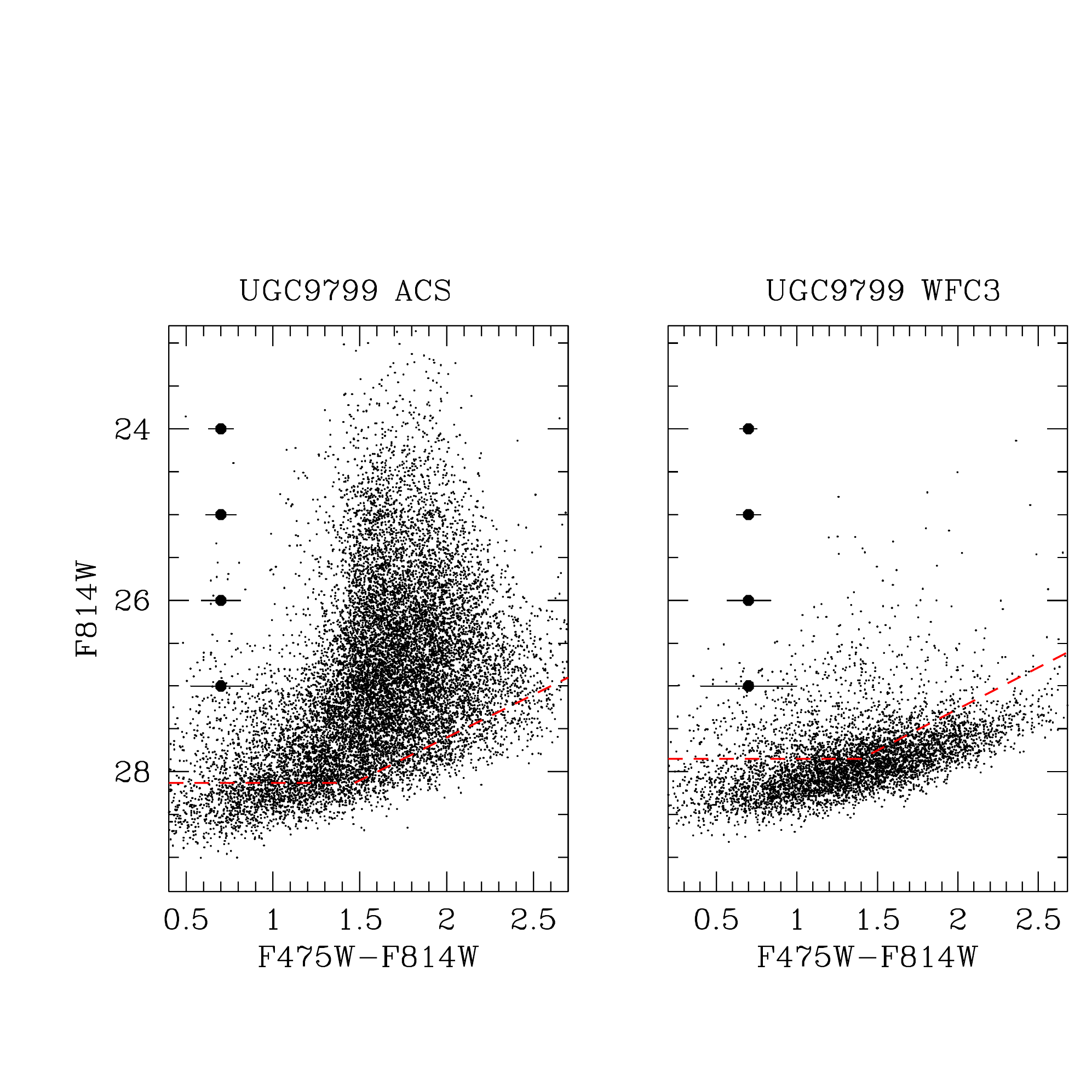}
\end{center}
\vspace{-1.0cm}
\caption{Color-magnitude diagrams for the measured starlike objects around
UGC 9799, for the ACS/WFC field (left panel) and the outlying WFC3/WFC field
(right panel).  
Here the native filter magnitudes $F475W, F814W$ are plotted,
closely equivalent to $I$ versus $(g'-I)$ in the Vegamag system.  These values are
not corrected for reddening.  For the WFC3
field, objects with $y < 95''$ are plotted to avoid contamination from another
galaxy (see text).  Detection completeness levels of $f=0.5$ are marked with 
the red dashed lines, and the photometric measurement uncertainties are indicated
by the errorbars at left.
}
\vspace{0.0cm}
\label{fig:ugc9799_cmd}
\end{figure*}

The color-magnitude diagrams for UGC 9799 are shown in Figure \ref{fig:ugc9799_cmd}.
The pattern seen in the CMD is reminiscent of what we found for NGC 6166
(Paper II), with a noticeable blue MP sequence centered near $(F475W-F814W) \simeq 1.6$
and a broader distribution of objects to the red, but with no clear `valley' at intermediate
colors.  In the WFC3 field (unlike for NGC 6166), remarkably few objects are seen in the
expected GC color range, suggesting that the `intragalactic' GC population 
within A2052 is small.  Note, however, that the data plotted here comprise
only the objects with $y < 95''$ in Fig.~\ref{fig:ugc9799_xy}.  We used only
the lower part of the WFC3 field to avoid the contamination
from the smaller galaxies in the top half of the field.  For WFC3, the large
number of objects fainter than the 50\% completeness line is a result of a very conservatively
faint initial detection threshold, so most of these `objects' are likely not to be real.
None of our analysis uses data fainter than the
completeness limit.

\begin{figure*}[t]
\vspace{-2.0cm}
\begin{center}
	\includegraphics[width=0.6\textwidth]{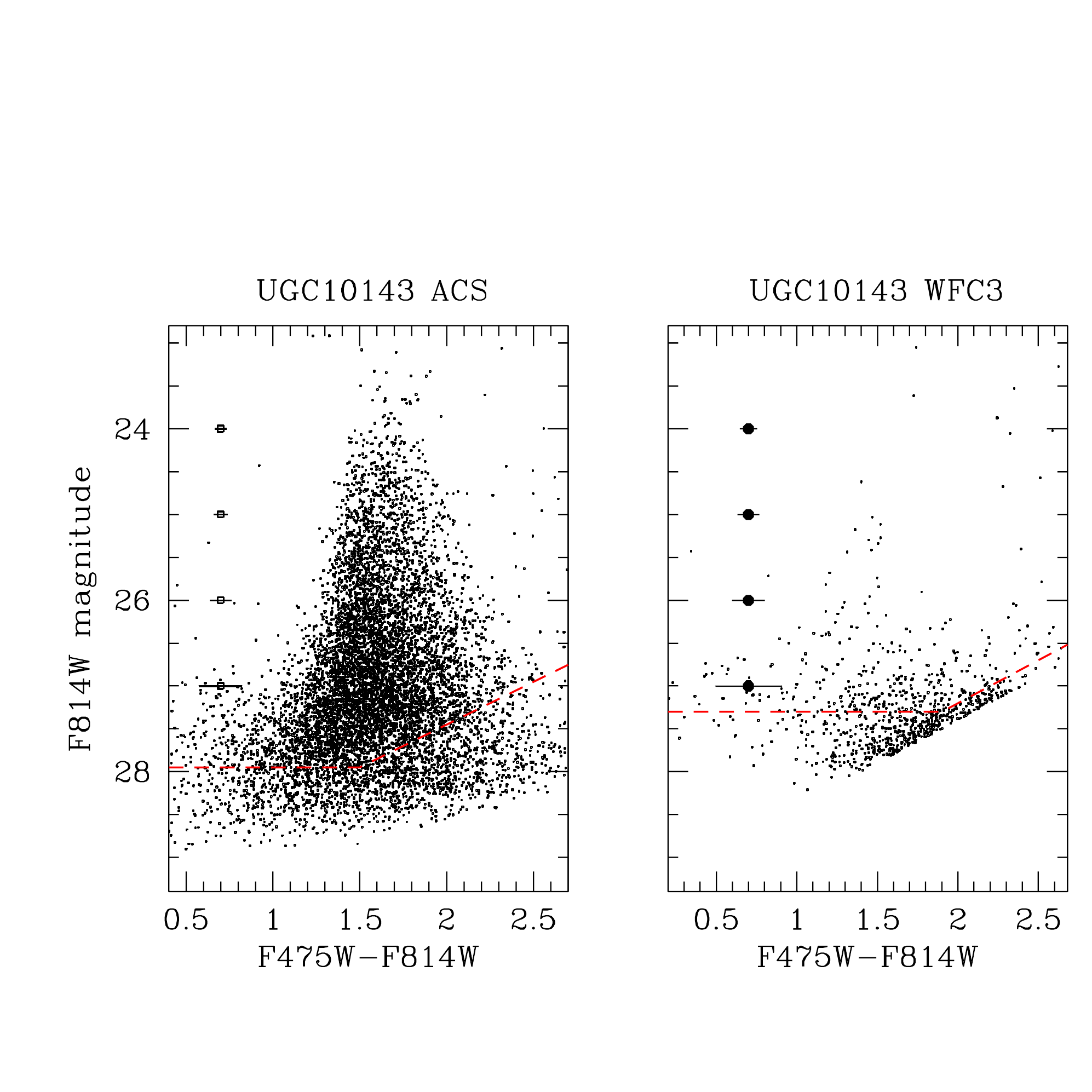}
\end{center}
\vspace{-1.0cm}
\caption{Color-magnitude diagrams for UGC 10143, for ACS (left) and
WFC3 (right).}
\vspace{0.0cm}
\label{fig:ugc10143_cmdpair}
\end{figure*}

\begin{figure}[t]
\vspace{0.0cm}
\begin{center}
\includegraphics[width=0.6\textwidth]{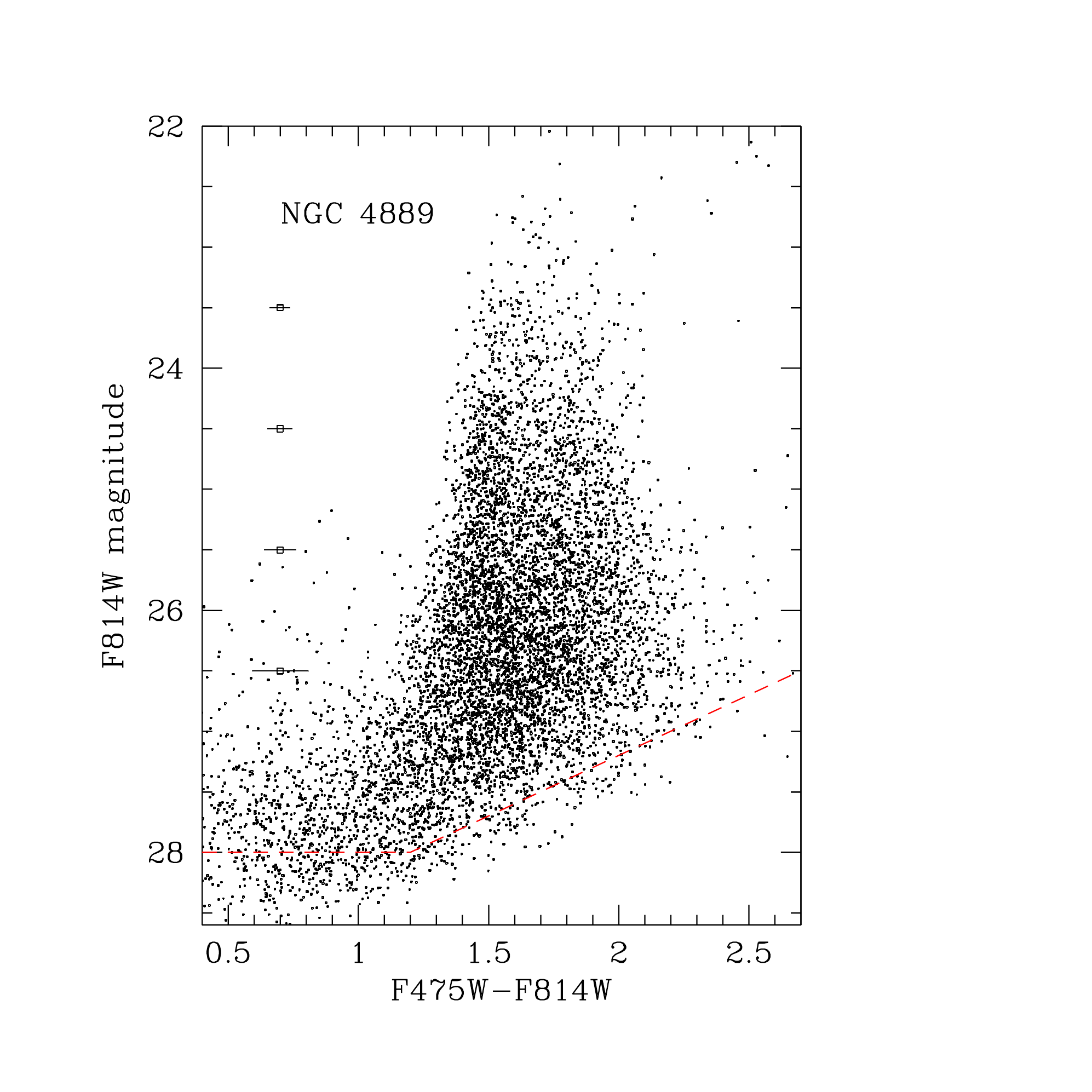}
\end{center}
\vspace{-0.5cm}
\caption{Color-magnitude diagram for 7892 measured starlike objects around
NGC 4889. Objects within $R < 15''$ of either NGC 4889 or NGC 4886 are excluded.}
\vspace{0.0cm}
\label{fig:ngc4889_cmd}
\end{figure}

\begin{figure*}[t]
\vspace{-3.0cm}
\begin{center}
\includegraphics[width=0.8\textwidth]{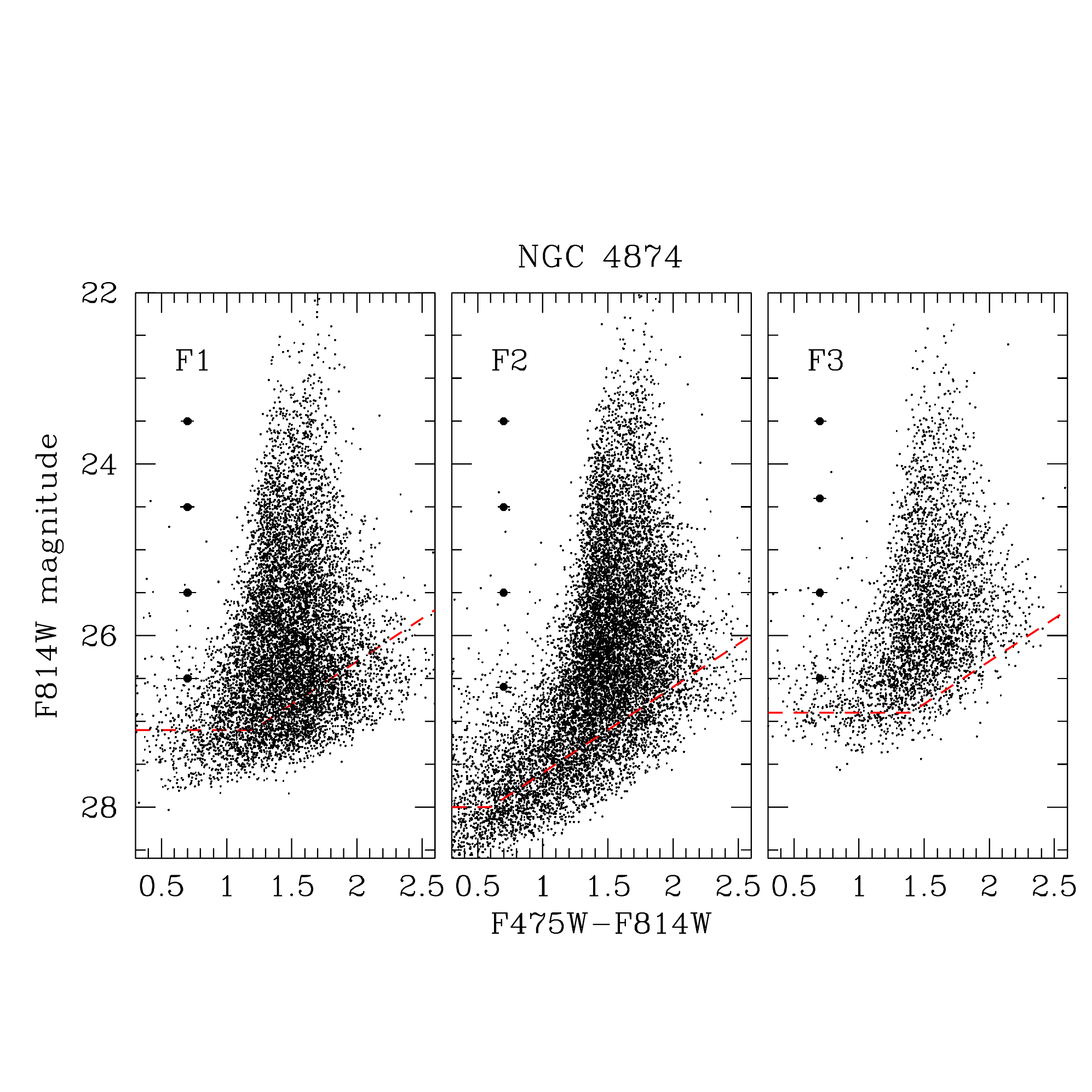}
\end{center}
\vspace{-1.5cm}
\caption{Color-magnitude diagrams for the measured starlike objects in the
three fields F1, F2, and F3 around NGC 4874. F1 and F3 each overlap partially with F2,
so the samples in these three diagrams are not fully independent. Note the differences
in limiting magnitude.}
\vspace{0.0cm}
\label{fig:ngc4874_cmd3}
\end{figure*}

\begin{figure}[t]
\vspace{0.0cm}
\begin{center}
\includegraphics[width=0.6\textwidth]{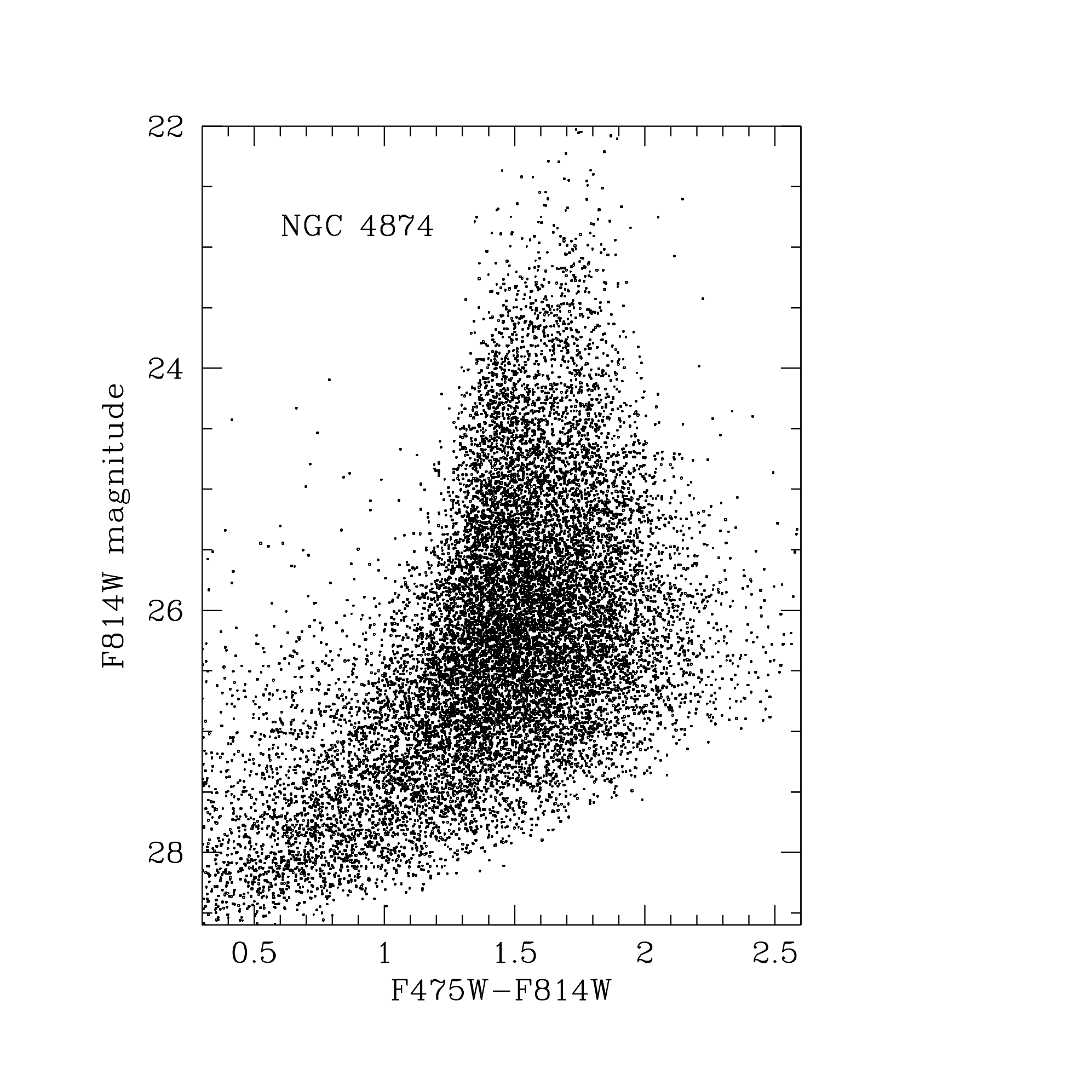}
\end{center}
\vspace{-0.5cm}
\caption{Color-magnitude diagram for the measured starlike objects around
NGC 4874, for the combination of three fields as described in the text.}
\vspace{0.0cm}
\label{fig:ngc4874_cmd}
\end{figure}

For UGC 10143 the ACS and WFC3 CMDs are shown in Figure \ref{fig:ugc10143_cmdpair}.
The overall color distribution resembles UGC 9799, although a closer look shows relatively
few red (MR) clusters; this will be quantified in the discussion below.  
Again, very few objects in the GC color range appear in the WFC3 field, where we have
excised only the datapoints within $15''$ of the small galaxy at upper left in
Fig.~\ref{fig:ugc10143_xy}.

For NGC 4889, the CMD is shown in Figure \ref{fig:ngc4889_cmd}.  In this case a more
noticeable classic separation between the MP and MR subpopulations is visible, primarily
because the MP sequence is narrower than for the other BCGs discussed here.  Still, significant overlap is present.  

For NGC 4874, the CMDs of the three ACS fields (F1, F2, F3 in Table \ref{tab:t})
are shown individually in Figure \ref{fig:ngc4874_cmd3}.  The F2 data are clearly the
deepest and a bit more internally precise than the shorter exposures of F1 and F3.  
The CMD for the combined fields is shown in Figure \ref{fig:ngc4874_cmd}. 
The net result leaves 16064 objects over all magnitudes, though in this combined graph
there is no single well defined limiting magnitude.  All the data are, however, highly
complete ($f > 0.9$) for $F814W \lesssim 26.0$.

The GC populations in these BCGs consist of many thousands of objects
and are completely dominant over any field contamination.  In Paper II we used photometry
from a local control field, similar data from the Hubble Ultra-Deep Field, and a standard model
for the expected population of Milky Way foreground stars to evaluate the field contamination
quantitatively.  The numbers
of contaminating objects within the magnitude and color range of interest here ($I \simeq 22-27$,
$(g-I) \simeq 1.2-2.5$) amount to less than a dozen starlike objects per ACS field and thus
are negligible by comparison with the GC populations.  In what follows, no corrections
are made for field contamination.

\section{The Color Distribution Functions}

\subsection{UGC 9799 and UGC 10143}

\begin{figure}[t]
\vspace{0.0cm}
\begin{center}
\includegraphics[width=0.5\textwidth]{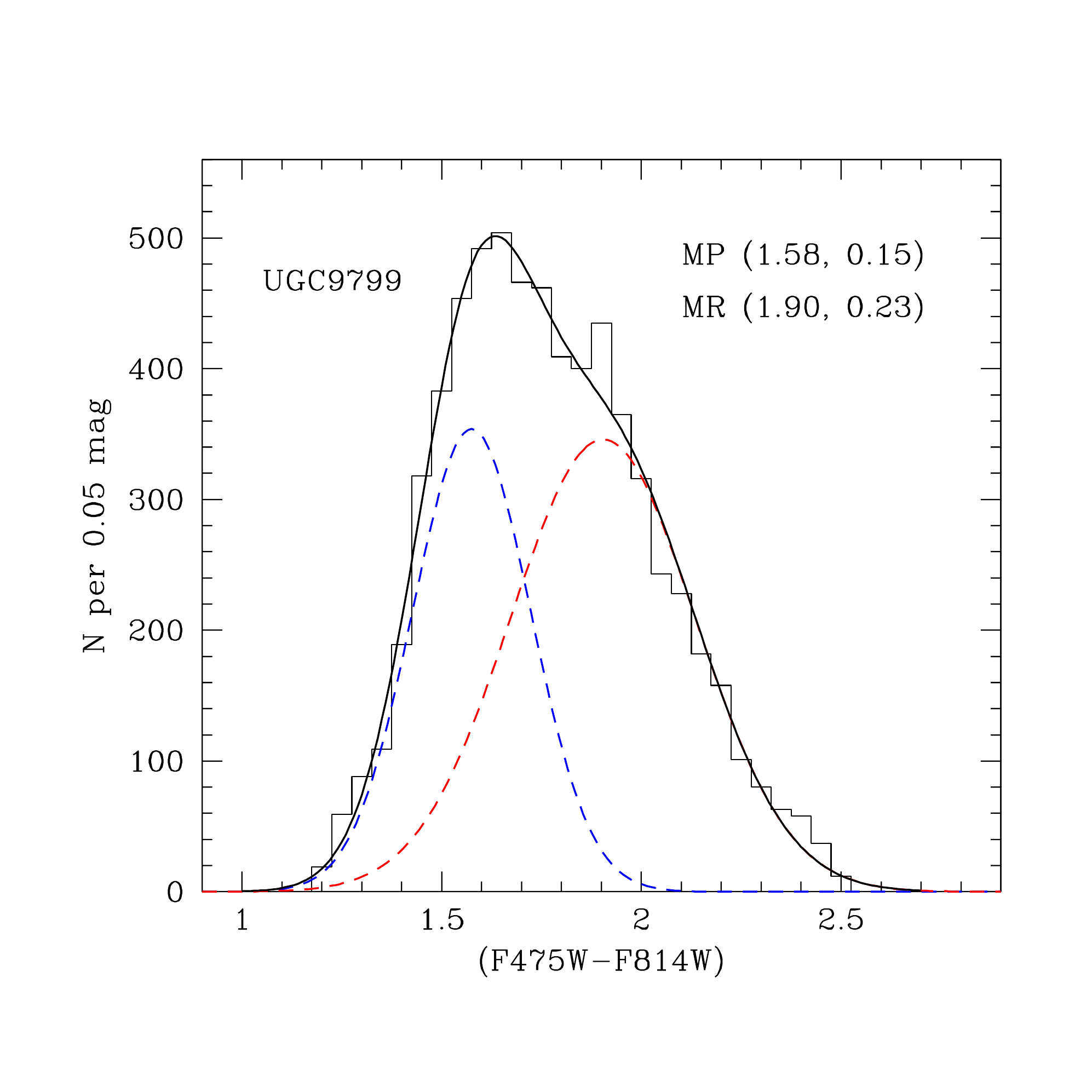}
\end{center}
\vspace{-0.5cm}
\caption{Color distribution function (CDF) for objects around UGC 9799
	with $F814W<27.0$ and $R > 10''$.  A bimodal-Gaussian fit to the
	data is shown by the superimposed curves; the numbers at upper right
give the means and standard deviations of the two modes.}
\vspace{0.0cm}
\label{fig:ugc9799_cdf}
\end{figure}

For UGC 9799, the CDF for 6630 objects brighter than $F814W = 27.0$, in the color range
$(F475W-F814W) = 1.2 - 2.5$, and $R > 10''$ from
galaxy center is shown in Figure \ref{fig:ugc9799_cdf}.  Quite clearly, the raw histogram
has a unimodal, skewed shape.  However, a bimodal-Gaussian model applied to the data
returns an excellent match to the entire histogram:  two modes are required, but more than
two are unnecessary.  As in Paper II, we use here the GMM
fitting code \citep{muratov_gnedin2010}.  In NGC 6166 we found the same pattern -- two 
broad and heavily overlapped CDF modes with no minimum or `valley' between them -- but UGC 9799 is even more extreme
(compare Fig.~11 from Paper II). Though the mode peaks $(\mu_1, \mu_2)$ are separated
by the same amount in color (0.32 mag, corresponding to $\simeq 0.8$ dex in metallicity), 
the dispersions of each mode are distinctly larger.
The $D-$statistic, a useful measure of the separation between modes relative to their dispersions
(see Muratov \& Gnedin and Paper II), is $D \simeq 1.70$, below the $D \gtrsim 2$ range    
where intrinsic bimodality can be strongly favored.

The fitting parameters for comparison are listed in Table \ref{tab:gmm}.
Here, the MP and MR modes have peak colors ($\mu_1, \mu_2$) and dispersions
($\sigma_1, \sigma_2$), $p_1$ is the fraction of objects belonging to the MP mode,
and $D$ measures the statistical significance of the mode separation.
For UGC 9799, the ACS field excludes the two satellite galaxies marked in 
fig.~\ref{fig:ugc9799_xy}, while
the WFC3 field excludes the upper part of the frame containing the smaller
galaxies as described above.

\begin{figure}[t]
\vspace{0.0cm}
\begin{center}
\includegraphics[width=0.5\textwidth]{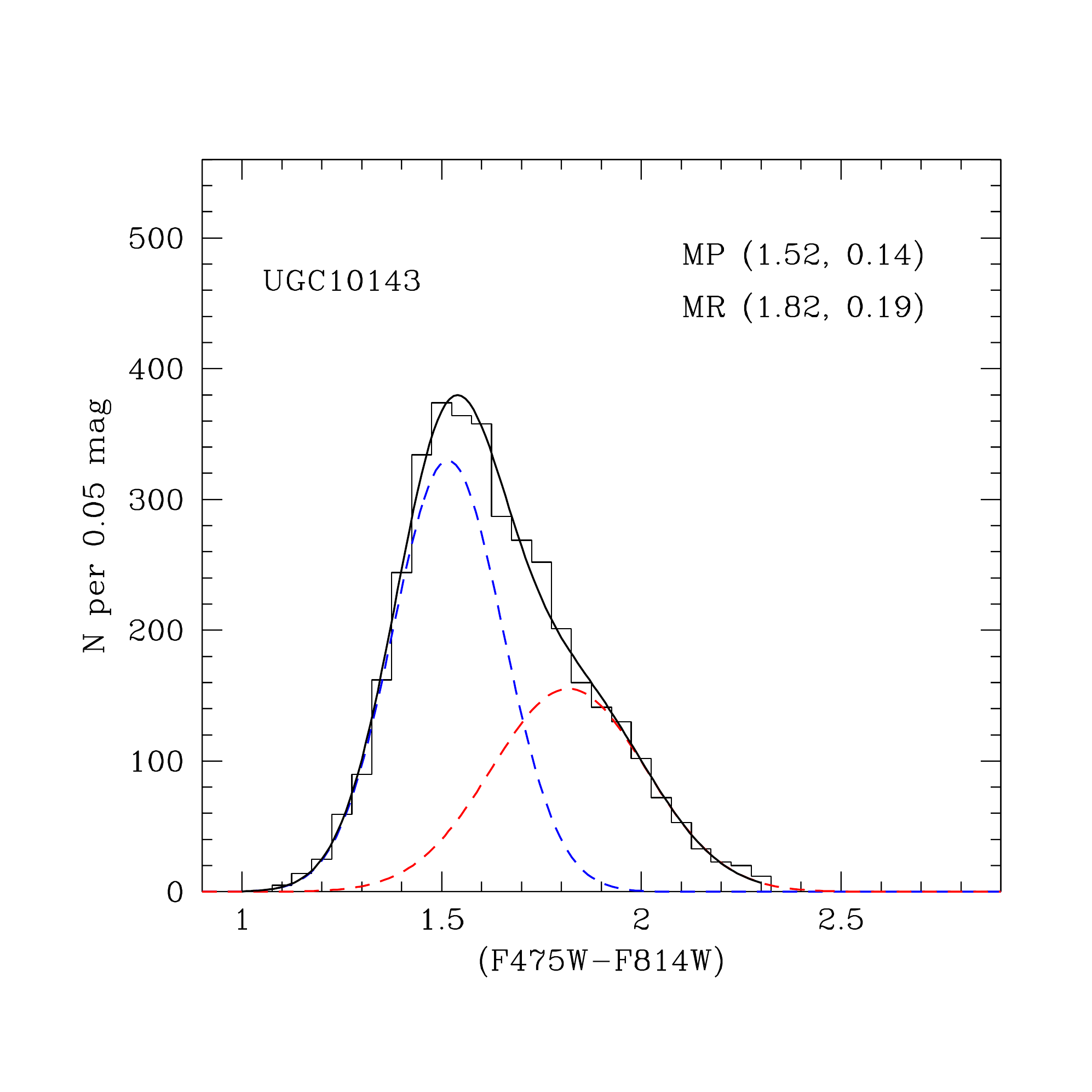}
\end{center}
\vspace{-0.5cm}
\caption{CDF for objects around UGC 10143
	with $F814W<27.0$ and $R > 20''$.  A bimodal-Gaussian fit and its parameters
	are shown by the superimposed curves and the numbers at upper right
as in the previous figure.}
\vspace{0.0cm}
\label{fig:ugc10143_cdf}
\end{figure}

For UGC 10143, the CDF for 3784 objects brighter than $F814W = 27.0$, in the color range
$(F475W-F814W) = 1.1 - 2.3$, and $R > 20''$ from
galaxy center is shown in Figure \ref{fig:ugc10143_cdf}.  The satellite galaxy at lower left 
in Fig.~\ref{fig:ugc10143_xy} is excluded.  The CDF 
has a unimodal, skewed shape as for UGC 9799, but again a bimodal-Gaussian model produces
an excellent fit with the results given in Table \ref{tab:gmm}. The main difference compared
with UGC 9799 is a noticeably lower proportion of the MR component (just 40\% of the total
GC population, compared with $\sim$60\% for NGC 6166 and UGC 9799).

\subsection{The Coma Giants}

\begin{figure}[t]
\vspace{0.0cm}
\begin{center}
\includegraphics[width=0.5\textwidth]{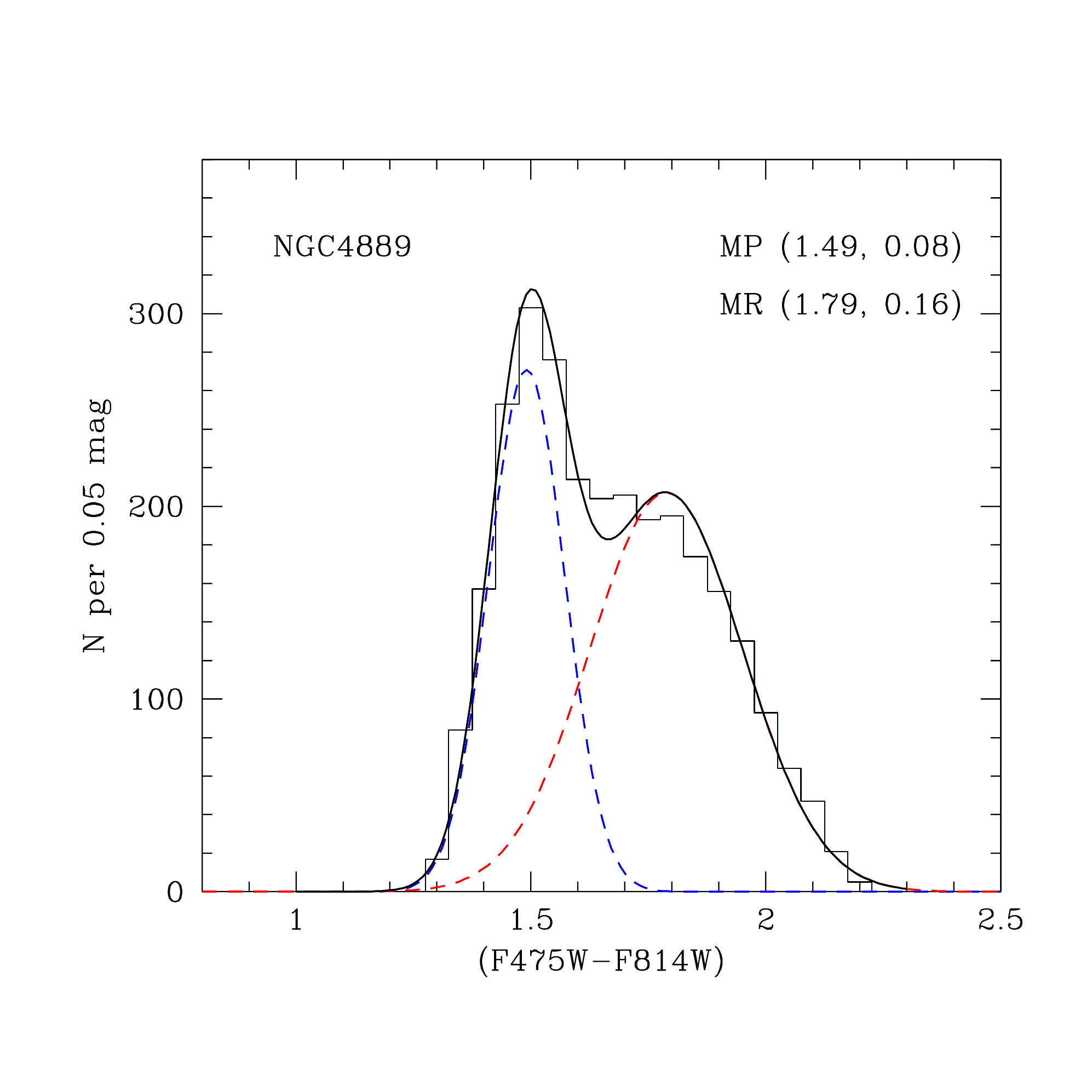}
\end{center}
\vspace{-0.5cm}
\caption{CDF for objects around NGC 4889 
	with $F814W = 23.5 - 26.0$ and $R > 15''$.  A bimodal-Gaussian fit and its parameters
	are shown by the superimposed curves and the numbers at upper right
as in the previous figure.}
\vspace{0.0cm}
\label{fig:ngc4889_cdf}
\end{figure}

For NGC 4889 the CDF for 2956 objects in the range $F814W = 23.5 - 26.0$ and 
$(F475W-F814W) = 1.3 - 2.2$ is shown in Figure \ref{fig:ngc4889_cdf}.  The innermost
circle $R < 20''$, most affected by background light and incompleteness, is not included.
The best-fit bimodal Gaussian shown in the figure does well at reproducing the wings of the
CDF and the blue peak, but less well in the intermediate color range $\sim 1.6-1.8$ where
noticeable discrepancies with the model occur.  Attempts at adding a third or fourth mode do not improve the
fit (the GMM solution damps down these additional modes to negligible levels).

\begin{figure}[t]
\vspace{0.0cm}
\begin{center}
\includegraphics[width=0.5\textwidth]{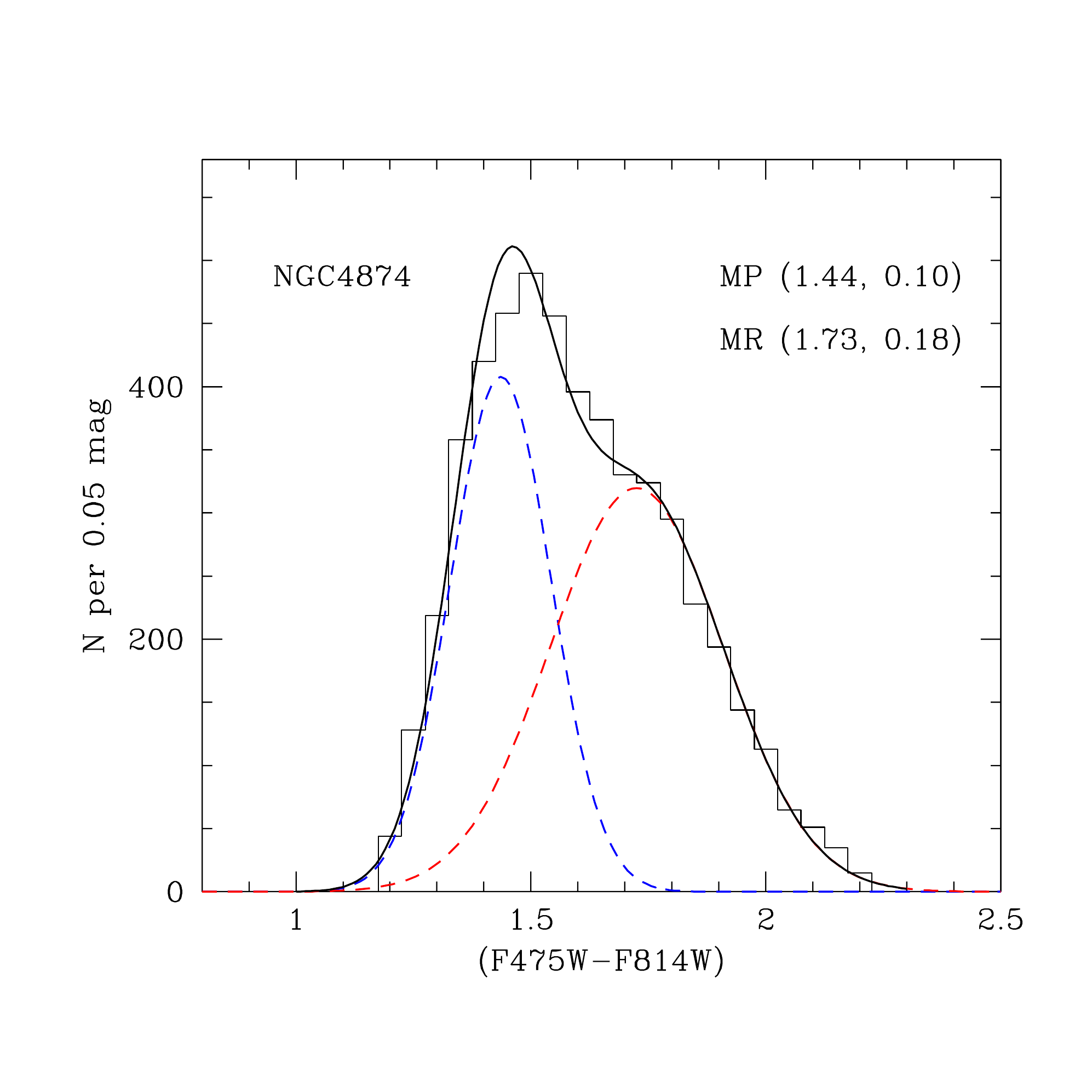}
\end{center}
\vspace{-0.5cm}
\caption{CDF for objects around NGC 4874 
with $F814W = 24.0-26.0$ and $R > 20''$, along with the best-fit bimodal Gaussian.} 
\vspace{0.0cm}
\label{fig:ngc4874_cdf}
\end{figure}

Lastly, for NGC 4874 we show the CDF for 5140 objects in the magnitude range $F814W = 24.0 - 26.0$ and
color range $(F475W-F814W) = 1.2 - 2.2$ in Figure \ref{fig:ngc4874_cdf}.  We restrict the
magnitude range at the bright end ($F814W < 24$) to deliberately avoid the high-luminosity
end of the GC distribution where the CDF becomes unimodal 
\citep[see][and the next section]{cho_etal2016}. We also exclude the range fainter than $\simeq 26$ 
where the Field 3 data start to 
become incomplete.  Fortunately, an adopted limit of $F814W \simeq 26.0-26.5$ for the Coma giants
corresponds to about the same luminosity ($M_I \simeq -9$) as in UGC 9799/10143, which are
about one magnitude more distant.  For NGC 4874, a bimodal-Gaussian fit produces
a match to the CDF, but again not as cleanly as in 
NGC 6166, UGC9799, or UGC10143:  the color range near the blue 
peak stands out as discrepant.  Nevertheless, for both the Coma giants
a standard Kolmogorov test (i.e. a one-sample KS test) does not suggest a statistically significant 
deviation from the bimodal model.
Just as for NGC 4889 above, adding more modes to the 
solution does not improve the overall fit, nor does
the imposition of equal variances (homoscedasticity).  

As we discussed extensively in Paper II, the internal precision of the color indices 
in the magnitude range of interest ($F814W \lesssim 27$) is high enough to be
easily capable of resolving the widths ($\sigma_1, \sigma_2$) of the two modes
without adding significant spreading to either one.  The evidence then suggests that
these high color dispersions seen in all of the BCGs, along with the near-linearity of the $(g-I)$ color
index, are due to the intrinsic metallicity spread of each mode.

\begin{table*}[t]
\begin{center}
\caption{\sc Bimodal Gaussian Fits}
\label{tab:gmm}
\begin{tabular}{llrllllcc}
\tableline\tableline\\
\multicolumn{1}{l}{Galaxy} &
\multicolumn{1}{l}{$F814W$ Range} &
\multicolumn{1}{r}{$n$} &
\multicolumn{1}{l}{$\mu_1 (\pm)$} &
\multicolumn{1}{l}{$\mu_2 (\pm)$} &
\multicolumn{1}{l}{$\sigma_1 (\pm)$} &
\multicolumn{1}{l}{$\sigma_2 (\pm)$} &
\multicolumn{1}{c}{$p_1$} &
\multicolumn{1}{c}{$D$}
\\[2mm] \tableline\\
NGC6166-ACS & 23.0-26.5 & 4712 & 1.401(0.009) & 1.719(0.015) & 0.122(0.005) & 0.178(0.006) & 0.42(0.04) & 2.08(0.12) \\
NGC6166-WFC3 & 23.0-26.5 &  147 & 1.324(0.021) & 1.674(0.074) & 0.136(0.033) & 0.244(0.013) & 0.71(0.12) & 1.77(0.95) \\
UGC9799-ACS & 20.0-27.0 & 6630 & 1.575(0.009) & 1.904(0.013) & 0.148(0.005) & 0.231(0.002) & 0.40(0.03) & 1.70(0.03) \\
UGC9799-WFC3 & 20.0-27.0 &  181 & 1.369(0.025) & 1.877(0.051) & 0.089(0.016) & 0.300(0.017) & 0.39(0.06) & 2.29(0.23) \\
UGC10143-ACS & 24.0-27.0 & 3784 & 1.516(0.011) & 1.818(0.033) & 0.138(0.006) & 0.193(0.012) & 0.60(0.07) & 1.80(0.19) \\
UGC10143-WFC3 & 24.0-27.0 &  119 & 1.473(0.093) & 1.921(0.092) & 0.205(0.048) & 0.083(0.048) & 0.77(0.20) & 2.87(0.31) \\
NGC 4889 & 23.5-26.0 & 2956 & 1.497(0.006) & 1.802(0.009) & 0.084(0.003) & 0.163(0.004) & 0.40(0.02) & 2.35(0.09) \\
NGC 4874 & 24.0-26.0 & 5140 & 1.436(0.006) & 1.725(0.009) & 0.107(0.003) & 0.184(0.003) & 0.43(0.03) & 1.92(0.05) \\
\\[2mm] \tableline
\end{tabular}
\end{center}
\vspace{0.4cm}
\end{table*}

\begin{figure}[t]
\vspace{0.0cm}
\begin{center}
\includegraphics[width=0.5\textwidth]{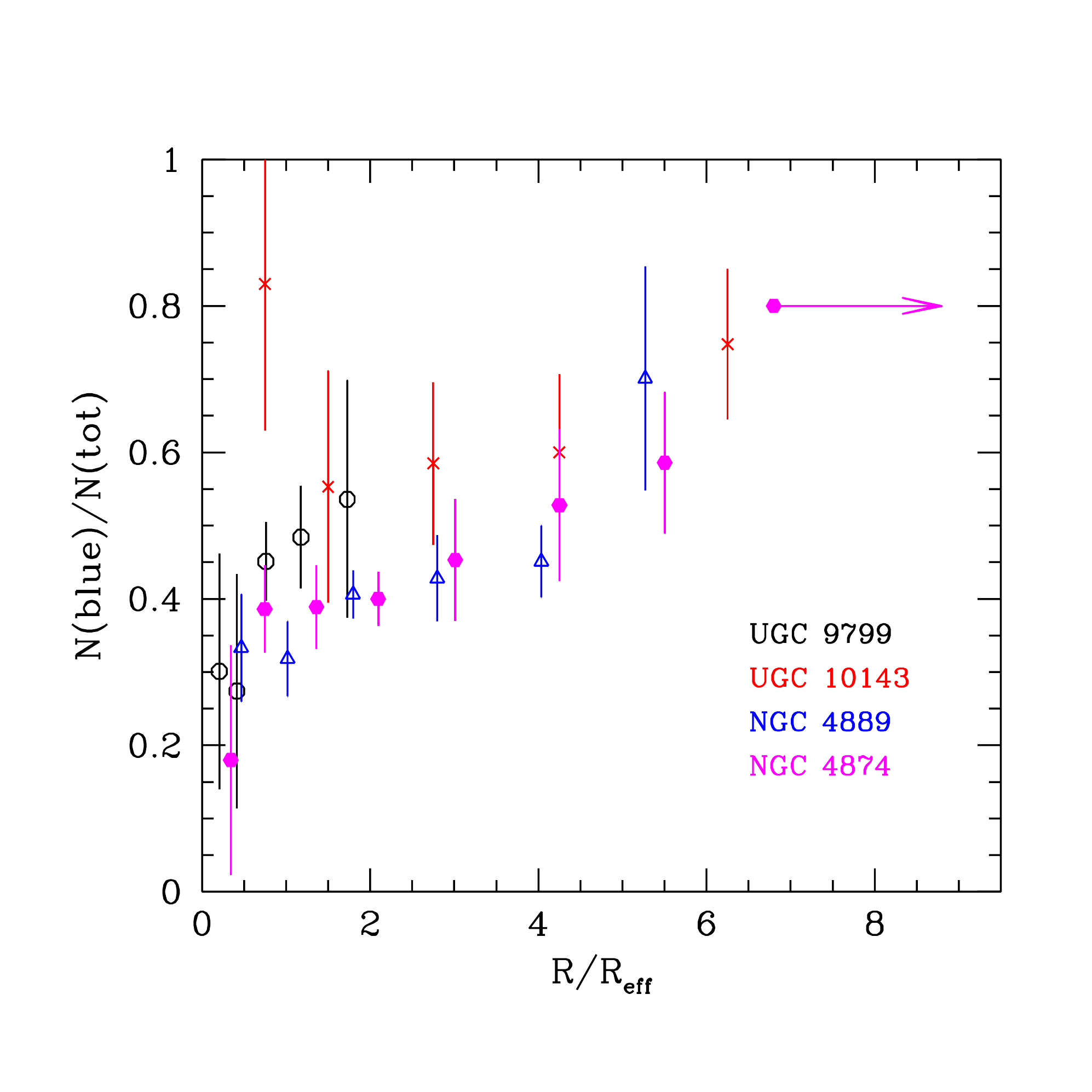}
\end{center}
\vspace{-0.5cm}
\caption{Blue fraction $p_1 = N(blue)/N(tot)$ as a function of projected
	galactocentric distance.  The outermost point for NGC 4874 (magenta
	arrow at upper right) refers to the GCs in the Intragalactic Medium
	of the Coma cluster.
}
\vspace{0.0cm}
\label{fig:fblue}
\end{figure}

\begin{figure*}[t]
\vspace{-1.2cm}
\begin{center}
\includegraphics[width=0.7\textwidth]{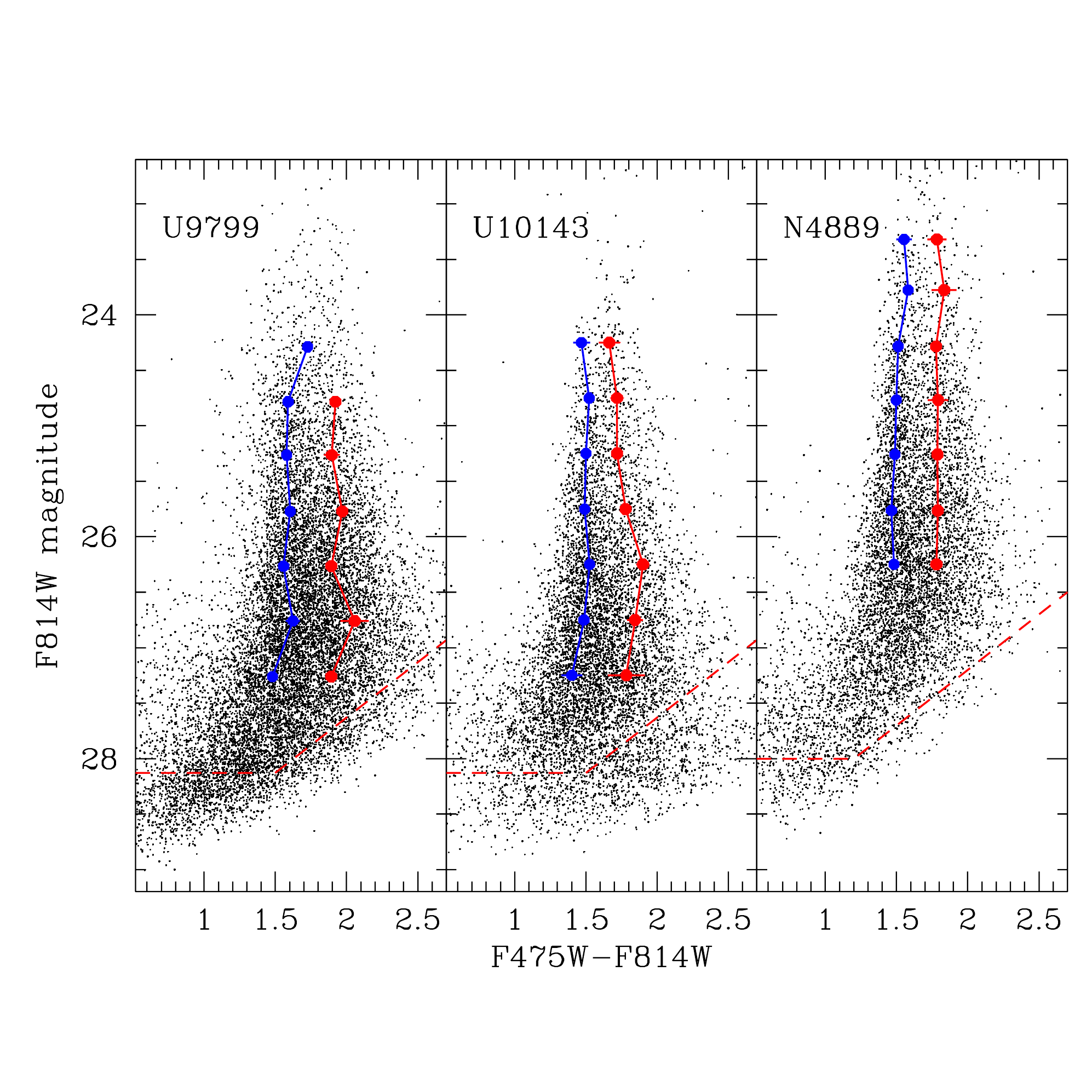}
\end{center}
\vspace{-1.0cm}
\caption{Mean points along the blue (MP) and red (MR) sequences for
UGC 9799, UGC 10143, and NGC 4889, calculated in 0.5-mag intervals.
In these three galaxies, little or no evidence appears for a systematic
trend of GC color with luminosity.}
\vspace{0.0cm}
\label{fig:mmr}
\end{figure*}

\subsection{MP and MR Populations Versus Radius:  Gradients?}

Previous evidence for many large galaxies (Paper II and references cited above) 
shows the common existence of radial metallicity gradients in their GC systems,
which show up primarily as a changing ratio of blue to red clusters with galactocentric
radius.  A convenient way to characterize these gradients is to plot the blue
fraction $p_1 \equiv N(blue)/N(total)$ versus $R_{gc}$.
For the four galaxies discussed here, the data for $p_1(R)$ are shown in Figure \ref{fig:fblue}.
In each case, GMM fits were done in radial zones, and the mean $\langle R \rangle$
for the clusters in each zone is
expressed as a ratio of the effective radius $R_{eff}$ of the galaxy light profile.

Although noticeable differences occur between galaxies in the overall mean level of
$p_1$, a repeated trend emerges for $p_1$ to rise fairly steeply from $R = 0$
out to $\simeq 1.5 R_{eff}$, then to plateau or rise more gradually until the blue fraction
reaches $p_1 \sim 0.5$, and finally 
beyond $\sim 4 R_{eff}$ to increase more steeply again.  For NGC 4874, we have
added an outermost data point from \citet{peng_etal2011}, who note that the 
intragalactic GC population becomes progressively more dominant beyond $R \simeq 270''$
($7 R_{eff}$) and that the ICL has $p_1 \simeq 0.8$.  The ICL value is shown as
in the Figure as an outward arrow.

\begin{figure*}[t]
\vspace{-0.1cm}
\begin{center}
\includegraphics[width=0.45\textwidth]{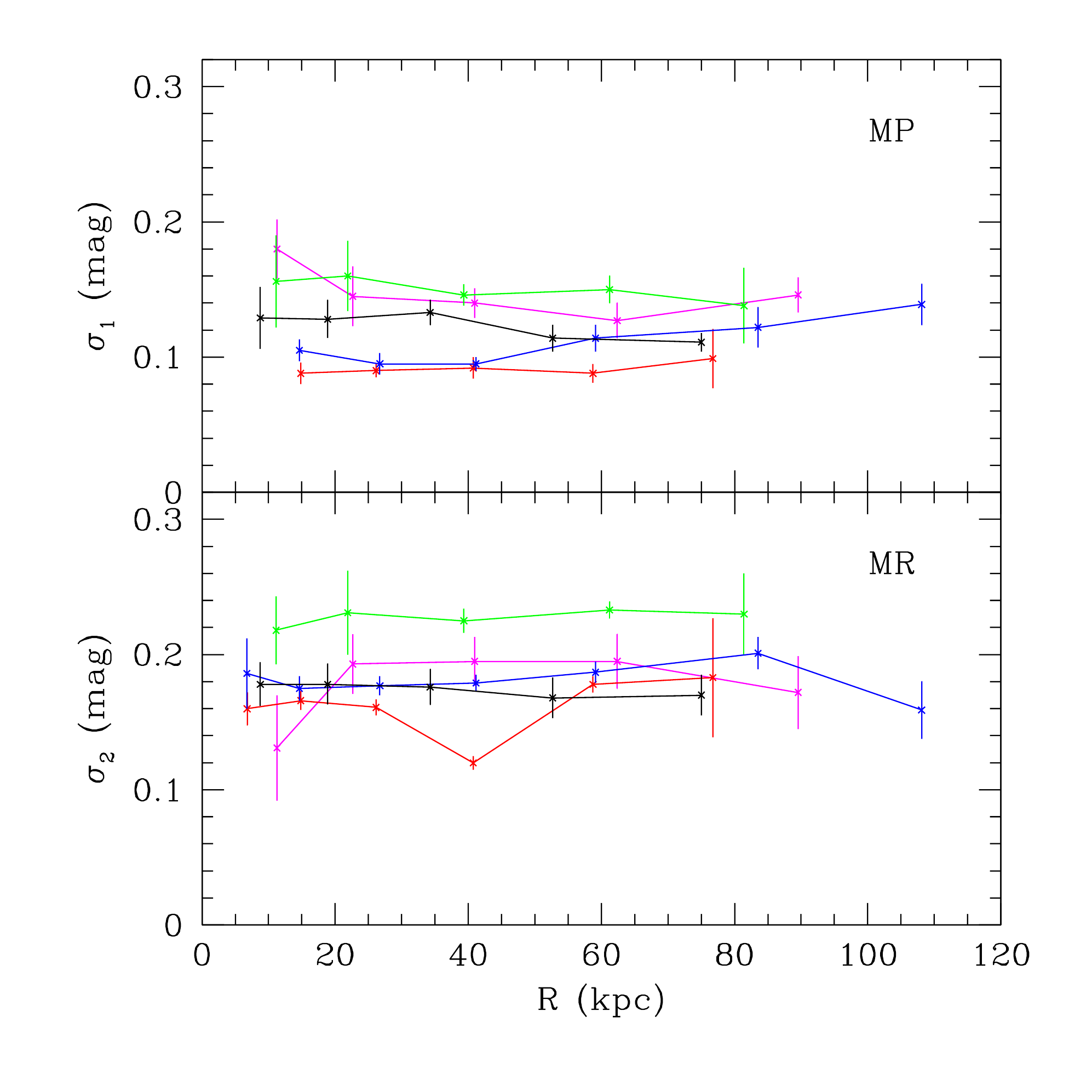}
\end{center}
\vspace{-1.0cm}
\caption{\emph{Upper panel:} Gaussian width $\sigma_1$ of the blue (MP) sequence
	for the five BCGs discussed in this paper, plotted as a function of radius, in kiloparsecs.  
	\emph{Lower panel:} Width of the red (MR) sequence $\sigma_2$ versus $R$.
	NGC 6166 is in \emph{black}, NGC 4874 in \emph{blue}, NGC 4889 in \emph{red},
UGC 9799 in \emph{green}, and UGC 10143 in \emph{magenta}. }
\vspace{0.0cm}
\label{fig:radsigma}
\end{figure*}

As emphasized above, a major result of this study is that the MDFs summed over all radii are broad with
heavy overlap between the standard MP and MR sequences.  Two straightforward ways this result
could be generated are that 
(a) the intrinsic dispersions of the two modes are actually small at a given radius, but there is a
strong radial gradient in the mean colors $\mu_1, \mu_2$ so that the two modes are blurred together
when all radii are combined; or alternately that
(b) the intrinsic dispersions $\sigma_1, \sigma_2$ are
high at all galactocentric radii, so that any smaller radial zone would be fairly representative of the
whole.  
In option (a), the MDF should actually appear more distinctly 
bimodal the more restricted the range of radii.

Option (b) appears to be the correct one. To test option (a) directly, we performed
GMM bimodal-fit solutions in smaller
radial bins.  These show that there is no significant variation of either $\mu_1$ or $\mu_2$ with radius (see
also Paper II for NGC 6166 in more detail).  A direct test of option (b) is in
Figure \ref{fig:radsigma}, where we show the dispersions $\sigma_1, \sigma_2$ versus mean radius,
and where $R$ is renormalized to the same physical scale (kiloparsecs) for all five galaxies discussed
here.  There are some differences in the mean $\sigma$'s from one galaxy to another, but the 
clear signal from this plot is that the two modes are both quite broad at all radii, and
furthermore that within any one GCS they undergo very little change with $R$.  In other words,
the MP and MR modes are heavily overlapped at all projected locations in their halos.
As expected, the most obviously bimodal system of the five (NGC 4889) has the lowest $\sigma-$values
in both modes.  (We recognize here, again, that for cases such as UGC 9799 and UGC 10143, 
where the MDFs are in fact quite smooth and continuous, discussing them in terms of two clearly
distinguishable `modes' begins to look like only a numerical exercise.)

The combined evidence points to a picture where the mean radial metallicity gradient in the
entire GCS is due entirely to a \emph{population gradient}, which shows up as a steadily increasing
ratio of MP to MR clusters with increasing radius.
Very similar results -- little change with radius in the mean colors and dispersions of both modes,
combined with a radial increase in $f(blue)$ -- have been found for other giant ellipticals including 
M87, M49, and M60 \citep{geisler_etal1996,lee_etal2008,harris2009b},
NGC 1399 \citep{bassino2006}, NGC 6861 \citep{escudero_etal2015}, NGC 4278 \citep{usher_etal2013},
and other BCGs and giants \citep{bassino_etal2008,harris2009a,faifer_etal2011}.  One notable exception is NGC 1407, which
shows distinct radial decreases in mean color \citep{forbes_etal2011}; small but nonzero gradients 
appear in intermediate-luminosity Fornax members \citep{liu_etal2011}.

\subsection{Mass-Metallicity Trends}

In NGC 6166 we found clear evidence for a mass-metallicity relation (MMR) along the
blue sequence in the sense that mean GC color becomes redder at higher luminosity.
Quantitatively the effect corresponded to a simple power law where heavy-element
abundance scales with GC mass as $Z \sim M^{\gamma}$ where $\gamma = {0.27 \pm 0.06}$ over almost the entire
luminosity range brighter than the GCLF turnover point.  A scaling similar to this
along the blue sequence has been found in several other giant and supergiant ellipticals at high statistical
significance \citep[e.g.][among others]{harris_etal2006,mieske_etal2006,strader_etal2006,wehner_etal2008,cockcroft_etal2009,peng_etal2009,faifer_etal2011,fensch_etal2014,cho_etal2016}.
No galaxy has revealed convincing nonzero color trends along the red MR sequence.  

For UGC 9799 and UGC 10143 the results from a similar analysis
are shown in Figure \ref{fig:mmr}.  For these two galaxies, the strong
overlap between the MP and MR components essentially continues upward to the brightest
magnitude ranges more or less unchanged, making clear conclusions about color slopes
quite difficult.  For UGC 9799, quantitatively we find
for the blue sequences $\gamma = 0.07 \pm 0.12$ if we use only the points
in the range $24.5 < F814W < 27.0$, but if we use all points $F814W < 27.5$ then
$\gamma = 0.48 \pm 0.21$.  For this galaxy, we cannot rule out any of $\gamma \sim 0$,
or a positive slope similar to other cases, or a nonlinear solution.
For UGC 10143, the mean points along the MP sequence give a much more consistent value
$\gamma = 0.01 \pm 0.05$, i.e. indistinguishable from zero.

In NGC 4889, which has more clearly separated blue and red modes, 
a more well defined trend along the blue sequence is seen, giving for a
linear fit $\gamma = 0.25 \pm 0.06$.  This is very similar to the
result we found for NGC 6166, $\gamma = 0.27 \pm 0.06$.

In NGC 4874, \citet{cho_etal2016} find strong evidence for a blue-sequence MMR,
and we will not repeat their extensive analysis here.  They find   
a nonlinear trend becoming steeper at brighter magnitudes, which is thus not well described
by a single slope $\gamma = const$.  Nevertheless, from our data using field F2 alone (the deepest and
most internally precise part of the photometry) we find $\gamma = 0.26 \pm 0.05$
using only a simple linear fit for $F814W < 26$.
For $F814W \lesssim 24$, the CDF becomes much
more nearly unimodal and symmetric; in this high-luminosity range, blue GCs become very rare while
the red sequence continues upward. 

\begin{figure}[t]
\vspace{-0.0cm}
\begin{center}
\includegraphics[width=0.4\textwidth]{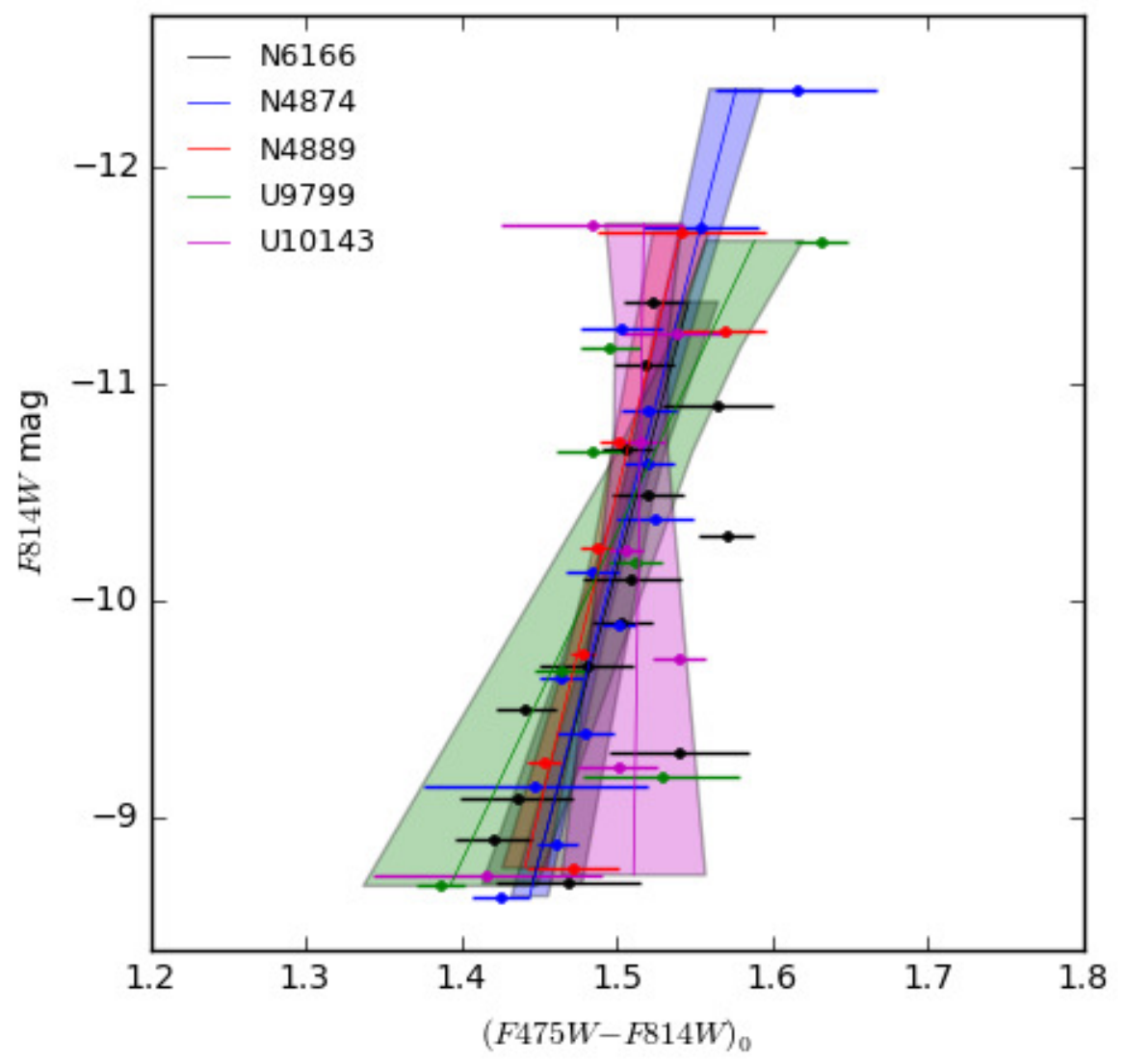}
\end{center}
\vspace{-0.0cm}
\caption{Mass-metallicity relations for the blue (metal-poor) sequences
in the five galaxies discussed here. For each galaxy the shaded region shows
the $\pm 1 \sigma$ uncertainty of the slope and intercept.  NGC 6166 is in grey,
NGC 4874 in blue, NGC 4889 in red, UGC 9799 in green, UGC 10143 in magenta.
Note the result for UGC 9799 
(light green region) is for a particular selection of the data points; see text.}
\vspace{0.0cm}
\label{fig:mmr_composite}
\end{figure}

A composite graph for the blue sequences in all five BCGs discussed here is shown in
Figure \ref{fig:mmr_composite}.  Here, the sequences have been shifted to a mean color
$(g-I)_0 = 1.5$ for all of them, to enable better direct comparison.  UGC 9799 is shown with
the solution from all the meanpoints, giving a positive but very uncertain slope.
NGC 4874, 4889, and 6166 are mutually very consistent at $\gamma \sim 0.25$, while UGC 10143
fairly clearly shows no color trend.

As more BCGs are added to the analysis, it is becoming clearer that the existence
of a MMR (or lack of one) is not universal, and even where one is present, no single
description (such as $\gamma = const$) may necessarily be valid.
UGC 9799 and UGC 10143 in Fig.~\ref{fig:mmr} can be added to NGC 4472 in Virgo, and NGC 1399 in Fornax,
as supergiant ellipticals that do not exhibit a definite MMR.
The model most often adopted to produce the blue-sequence trend is that of \citet{bailin_harris2009},
which is based on the approach that self-enrichment of the cluster stars 
takes place while the GC is still in its formation
stages and thus while both gas and young stars are mixed together. Quantitative 
examples of fits of this self-enrichment model to various data are in
\citet{mieske_etal2010,cockcroft_etal2009,forbes_etal2010,harris_etal2010,fensch_etal2014}.
It is worth emphasizing that the use of BCGs in particular for model tests
is crucial, because only the BCGs have large enough numbers of GCs to populate the
highest-luminosity range $L \gtrsim 10^6 L_{\odot}$ where the color trend becomes most obvious.

In Paper II we suggested that the Bailin/Harris model, at least in its basic form, has difficulty
matching the observed range of MMRs (essentially, the observed range of blue-sequence slopes $\gamma$).
If the MMR is due entirely to self-enrichment then it should be driven by 
very local conditions for the structure of the proto-GC, primarily its mass and 
central concentration (scale radius).  The higher the mass and the smaller 
the scale radius, the more efficiently gas is retained in the protocluster and 
the more effect pre-enrichment can have.  But this scenario then
suggests that the MMR should look similar from one galaxy to the next, with no immediately obvious role
for larger-scale environmental effects that could differ strongly between host galaxies.
In addition, self-enrichment in this model is highly nonlinear in cluster 
mass and thus it would predict that the slope $\gamma$ should 
increase with GC mass, which is not observed in all cases.
Another and more physically based obstacle faced by internal self-enrichment is simply that
it would require a fairly extended initial star formation period of $\gtrsim$10 My 
for the first round of SNeII to appear and contaminate the remaining gas in the protocluster,
which could then go on to form more low-mass stars of higher metallicity \citep[see][]{bailin_harris2009}.
But direct modelling of GC formation within giant molecular clouds indicates that most
of the star formation happens within $\lesssim 4$ My \citep[e.g.][]{howard_etal2016,hartmann_etal2012}.
Observations of young clusters indicate small internal age ranges as well
\citep[e.g.][]{lada_lada2003,schneider_etal2014,melena_etal2008,andersen_etal2009}.  
The high-mass regime suitable for GCs has, however, not yet been probed either in theory or
observation:  ideally we would
like to determine the expected internal age range for young clusters 
at masses well above $10^5 M_{\odot}$.  It is potentially promising that in 
the recent \citet{li_etal2016} models, cluster
formation proceeds over $\gtrsim 15$ My in some of the most massive cases.

An alternate approach, though at this point still simplistic, would be to introduce \emph{pre-}enrichment 
of the proto-GCs \citep{forte_etal2007,vandalfsen_harris2004} and invoke higher pre-enrichment
for higher-mass clusters.  Different amounts of pre-enrichment 
among GCs, presumably drawn from their host giant molecular clouds, 
would in principle allow for a wider range of environmental influences.  It is not yet
clear, however, exactly how pre-enrichment should depend on GC mass.

BCGs are likely to be products of mergers, and the sheer number of mergers could also have
differed noticeably from one case to another.  If we then assume that 
their progenitors experienced different
levels of pre-enrichment before the formation of their MP clusters, then in general
the MP sequence in the final combined BCG should have a higher internal color dispersion
and a weaker net MMR slope in the cases where more mergers took place.
In observational terms, $\sigma_1$ should increase as $\gamma$ decreases.  
Some hints that this is the case can be seen from the comparison of UGC 9799 and UGC 10143 
with the Coma giants, from the numbers in Table \ref{tab:gmm}.  

It also remains unclear whether or not the MMR phenomenon is connected with
the multiple stellar populations that have been detected within massive GCs in the Milky Way, an issue that
continues to be a serious challenge for modelling \citep[e.g.][]{renzini_etal2015}.
For additional discussion, see Paper II and \citet{fensch_etal2014}.

\section{Spatial Distributions}

\subsection{Radial Profiles}

\begin{figure}[t]
\vspace{0.0cm}
\begin{center}
\includegraphics[width=0.5\textwidth]{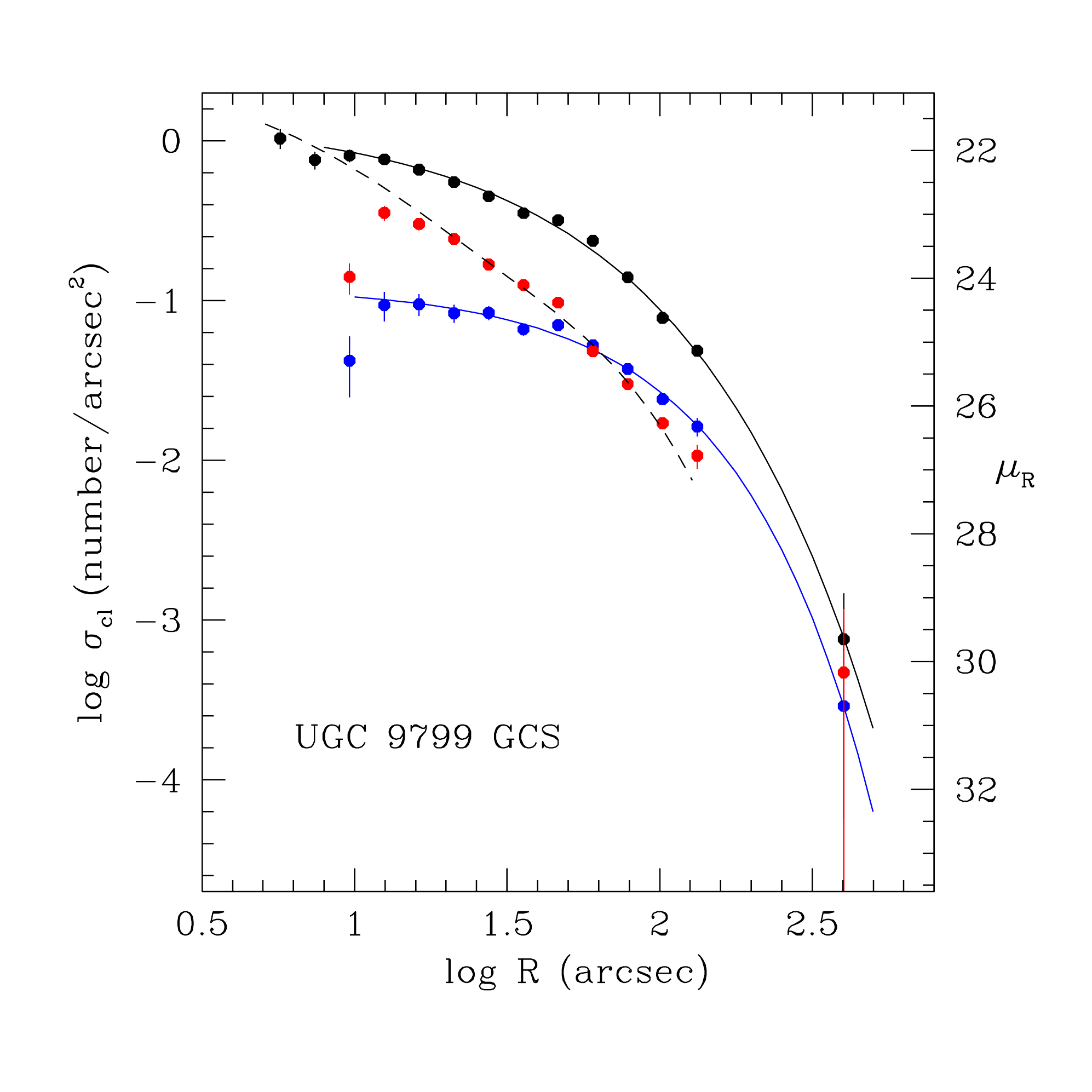}
\end{center}
\vspace{-0.5cm}
\caption{UGC 9799: Radial profile for the entire GC system (black points and solid curve),
	the EMP clusters (blue points and curve), and the EMR clusters (red points
	and curve).  The dashed line fitted to the EMR datapoints shows the integrated
	surface brightness profile of the galaxy's halo light in $\mu_R$ \citep{seigar_etal2007}
	with scale shown along the right-hand axis (see text).
}
\vspace{0.0cm}
\label{fig:ugc9799_radprof}
\end{figure}

\begin{figure}[t]
\vspace{0.0cm}
\begin{center}
\includegraphics[width=0.5\textwidth]{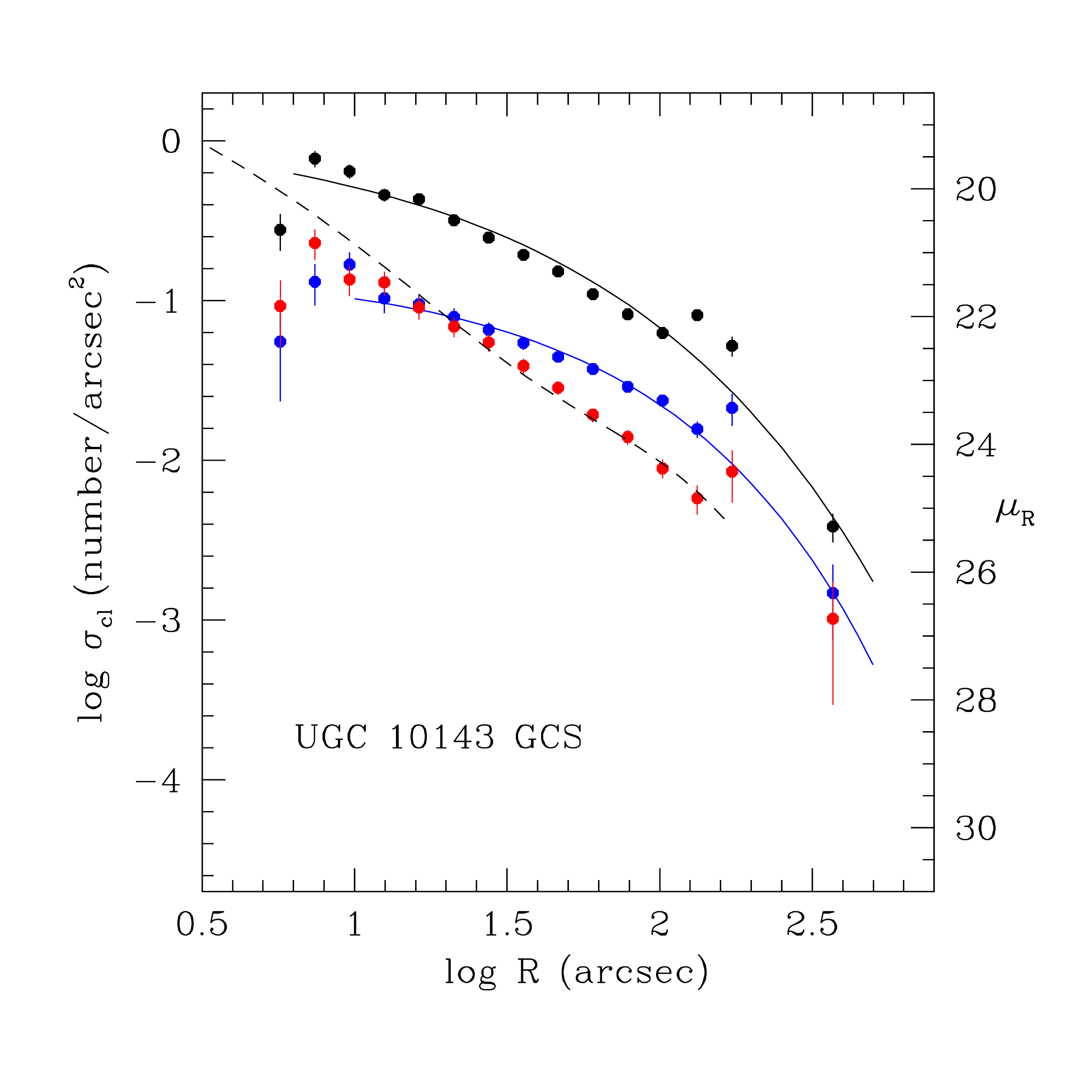}
\end{center}
\vspace{-0.5cm}
\caption{UGC 10143: Radial profile for the entire GC system,
	the EMP clusters, and the EMR clusters, with symbols as in the previous figure.
	The S\'ersic function fit is shown for the entire system (black line) and
	the EMP clusters (blue line).
	The dashed line fitted to the EMR datapoints shows the integrated
	surface brightness profile $\mu_R$ of the galaxy's halo light,
	from \citet{donzelli_etal2011}.
}
\vspace{0.0cm}
\label{fig:ugc10143_radprof}
\end{figure}

\begin{figure}[t]
\vspace{0.0cm}
\begin{center}
\includegraphics[width=0.5\textwidth]{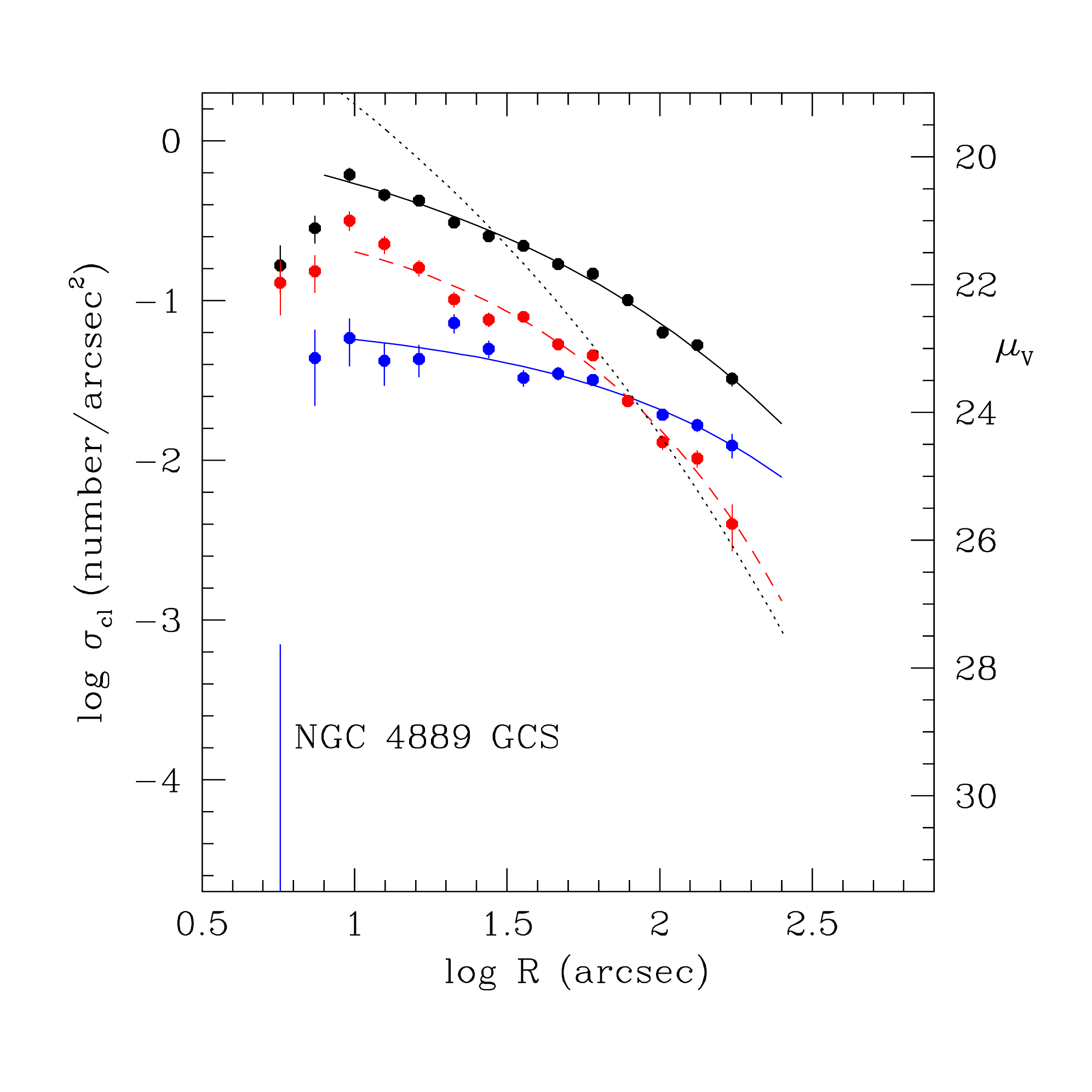}
\end{center}
\vspace{-0.5cm}
\caption{NGC 4889: Radial profile for the entire GC system,
	the EMP clusters, and the EMR clusters, with symbols as in the previous figure.
	The S\'ersic fits to each of the three components are shown as the solid line
	(all GCs), dashed blue line (EMP), and red line (EMR).  The dotted line
	indicates the integrated $V-$band surface brightness profile $\mu_V$ of the halo light
	from \citet{pahre1999}.
}
\vspace{0.0cm}
\label{fig:ngc4889_radprof}
\end{figure}

At the distance of these target galaxies the ACS field of view is large enough to enclose
a significant radial range of their halos.  In Figures \ref{fig:ugc9799_radprof}, 
\ref{fig:ugc10143_radprof}, and \ref{fig:ngc4889_radprof} 
the projected radial distributions $\sigma_{cl}$ (number of GCs per
arcsec$^2$) of the GCs are shown for UGC 9799, UGC 10143, and NGC 4889,
along with fits to S\'ersic-type functions obtained by $\chi^2-$minimization,
\begin{equation}
	 \sigma_{cl} \, = \, \sigma_e {\rm exp}(-b_n [({R \over R_e})^{1/n} - 1] ) \, 
 \end{equation}
or else to simple power-law form where appropriate.
In the first two figures, the outermost datapoints are from the outlying WFC3 field, which allow us to
track $\sigma_{cl}$ out to nearly $R \simeq 300$ kpc.
We note that the radial distributions for the GCs in NGC 4874 are analyzed by \citet{peng_etal2011} and
\citet{cho_etal2016} and we do not repeat their discussion here. 
They find clearly that the MP clusters follow a distinctly
shallower distribution than the MR clusters.

\begin{figure*}[t]
\vspace{0.0cm}
\begin{center}
\includegraphics[width=0.7\textwidth]{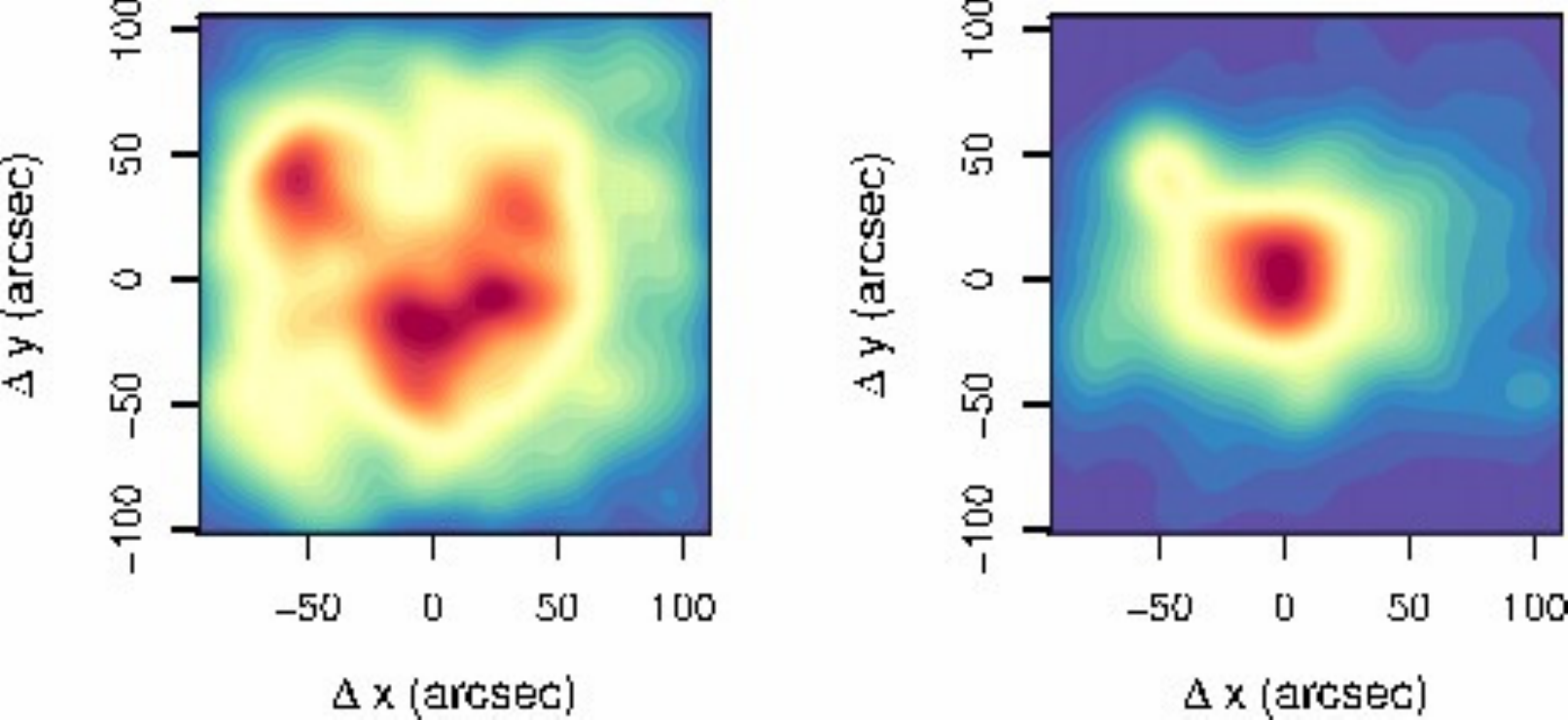}
\end{center}
\vspace{-0.5cm}
\caption{Spatial distributions in UGC 9799 for the extreme-MP clusters (left panel)
and the extreme-MR clusters (right panel), shown as smoothed isocontour maps.}
\vspace{0.0cm}
\label{fig:ugc9799_xy2}    
\end{figure*}

\begin{figure*}[t]
\vspace{-0.0cm}
\begin{center}
\includegraphics[width=0.7\textwidth]{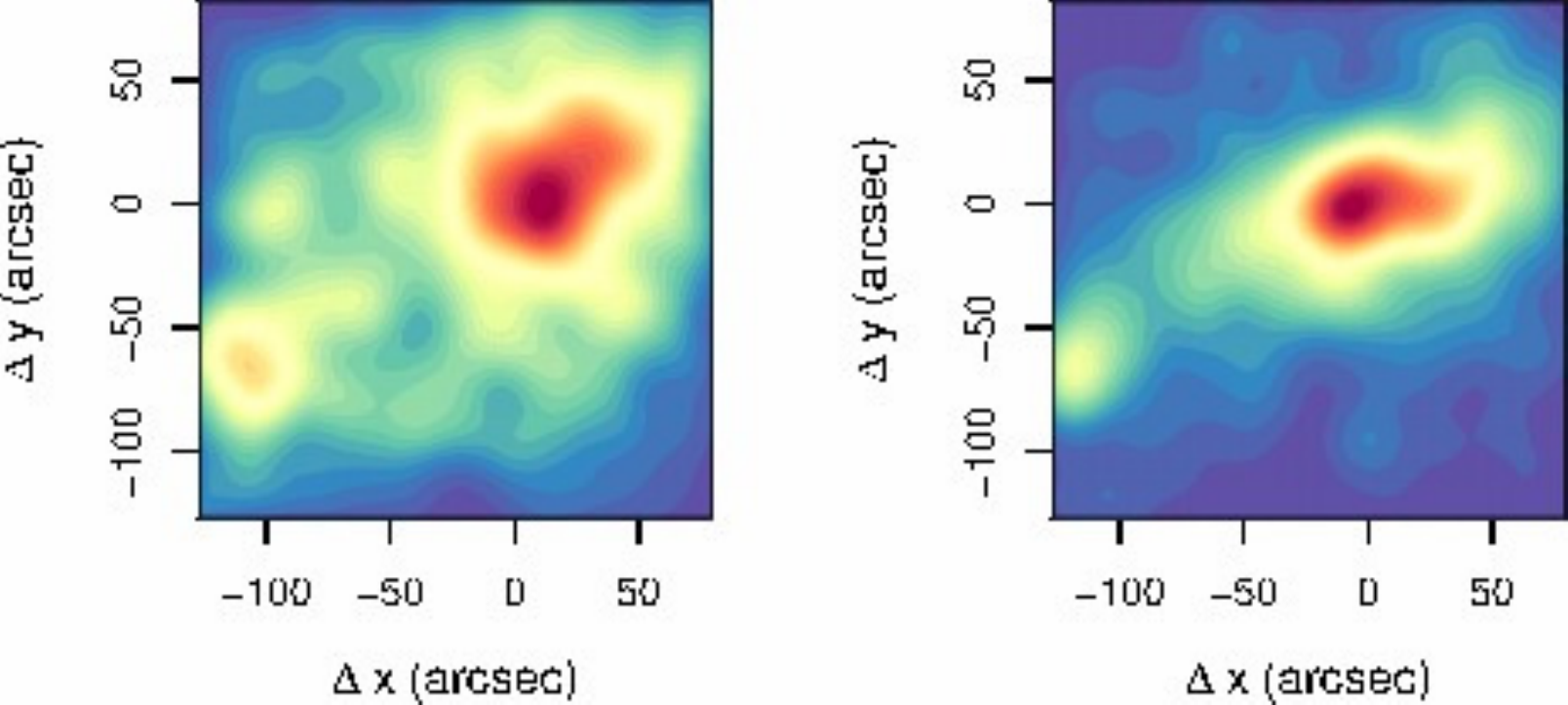}
\end{center}
\vspace{-0.5cm}
\caption{Spatial distributions in UGC 10143 for the EMP clusters (left panel)
and the EMR clusters (right panel).  }
\vspace{0.0cm}
\label{fig:ugc10143_xy2}    
\end{figure*}

\begin{figure*}[t]
\vspace{-0.0cm}
\begin{center}
\includegraphics[width=0.7\textwidth]{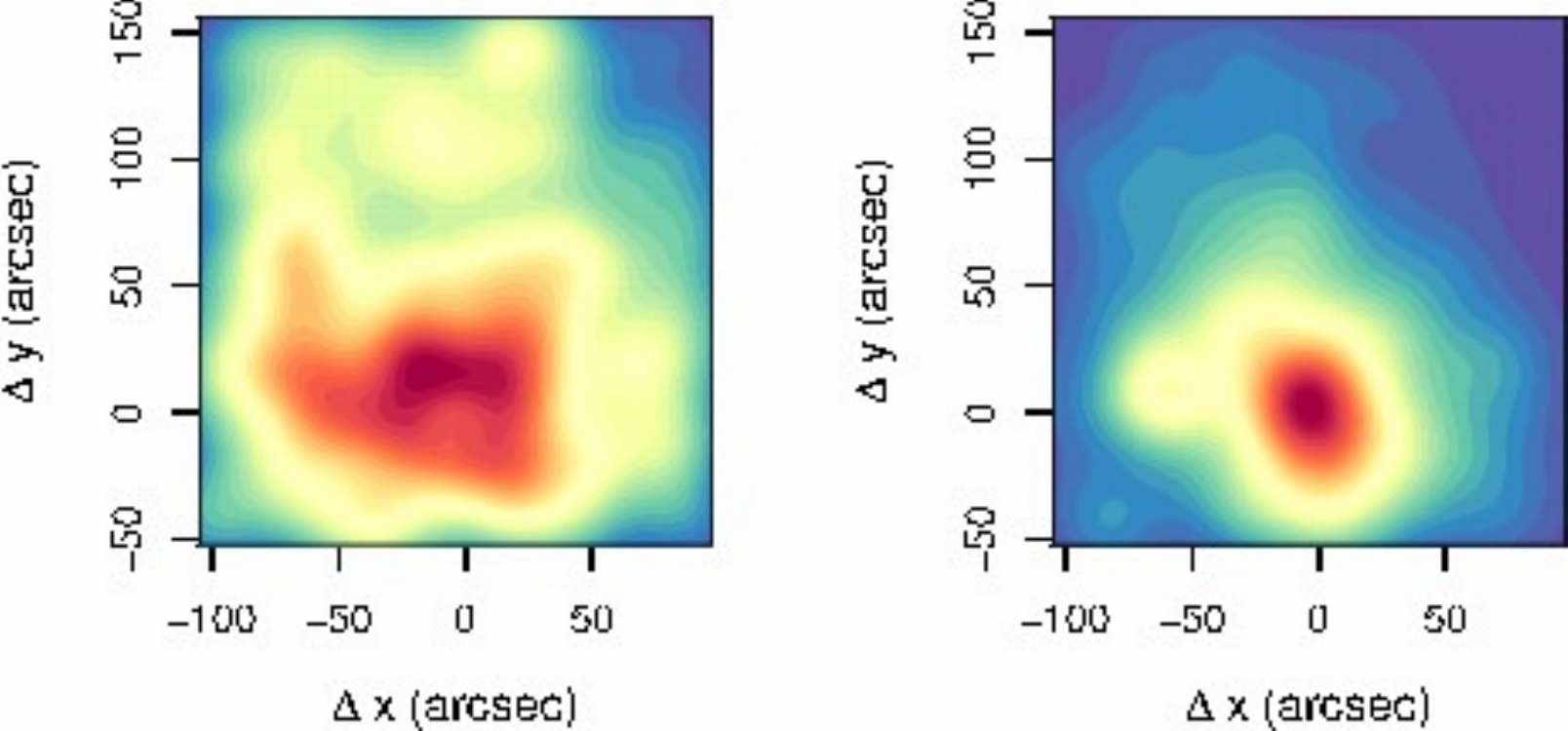}
\end{center}
\vspace{-0.5cm}
\caption{Spatial distributions in NGC 4889 for the EMP clusters (left panel)
and the EMR clusters (right panel).  }
\vspace{0.0cm}
\label{fig:ngc4889_xy2}    
\end{figure*}

\begin{figure*}[t]
	\vspace{-0.0cm}
\begin{center}
\includegraphics[width=0.7\textwidth]{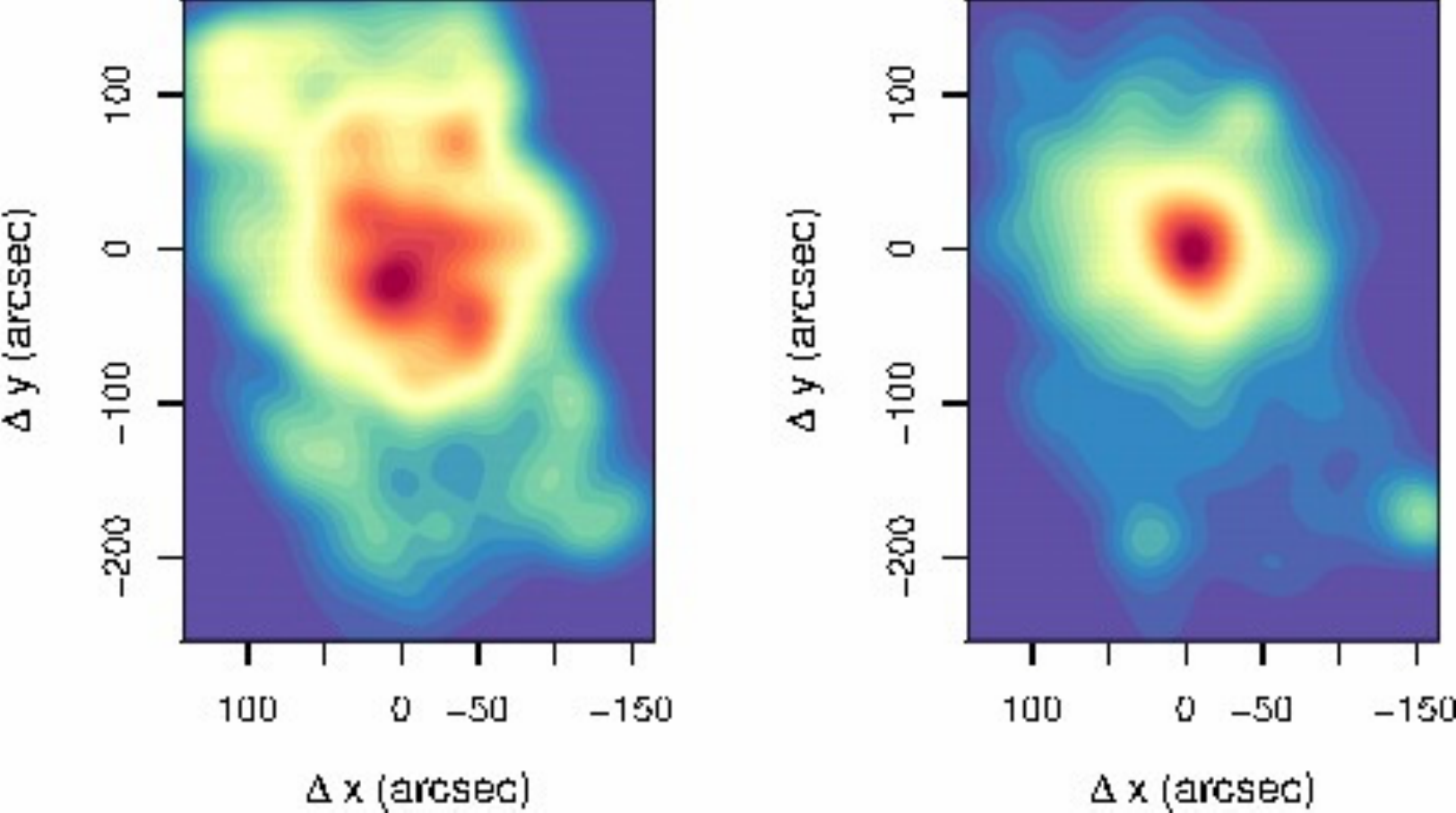}
\end{center}
\vspace{-0.5cm}
\caption{Spatial distributions in NGC 4874 for the EMP clusters (left panel)
and the EMR clusters (right panel).  }
\vspace{0.0cm}
\label{fig:ngc4874_xy2}    
\end{figure*}

The fact that the blue/red GC fraction increases with radius also means the
radial profiles are a function of metallicity.  
To minimize the effects of the strong overlap between the MP and MR modes,
we follow Paper II and define the \emph{extreme metal-poor} (EMP) clusters as those bluer than
the blue-mode peak, and \emph{extreme metal-rich} (EMR) clusters as those redder than the red-mode
peak.  Although this step eliminates half the total GC sample, it gives a clearer view
of the structural differences versus metallicity.  
In Figures \ref{fig:ugc9799_xy2} -- \ref{fig:ngc4874_xy2}
the $xy$ distributions for the EMP and EMR clusters in each galaxy
are shown in the form of smoothed isocontour maps.  The differences in central concentration 
are evident, with the EMP component distributed much more widely and often less symmetrically
than the EMR clusters.

\smallskip \noindent \emph{UGC 9799:} The profile fits use the data $F814W < 27.0, R > 10''$.
The solutions for the radial fits are summarized in Table \ref{tab:radial}.
The EMR component does not fit a 
single S\'ersic profile as well, but it is certainly more centrally concentrated than the
EMP component:  very roughly, in simple power-law terms for $R \gtrsim 30''$ the EMP clusters
follow $\sigma \sim R^{-1.4}$ and the EMR clusters $\sigma \sim R^{-2.1}$.  For comparison,
in NGC 6166 we found $\sigma(EMP) \sim R^{-1.0}$, $\sigma(EMR) \sim R^{-1.8}$.  Thus the overall
halo of UGC 9799 is a bit more centrally concentrated, but the difference between
EMP and EMR components is similar.  Perhaps more importantly, the EMR distribution matches
the surface brightness of the halo light well, as shown by the dashed line in Fig.~\ref{fig:ugc9799_radprof}.
Surface-brightness profiles in the $R$ filter have been measured by
\citet{seigar_etal2007} with a double S\'ersic profile out to $R \simeq 150''$, whereas 
\citet{donzelli_etal2011} measure it out to $R \simeq 100''$ and fit it to a single S\'ersic profile.
Here we adopt the Seigar et al.~data.  

\smallskip \noindent \emph{UGC 10143:}  Again we use all data $F814W < 27.0, R> 10''$ for profile fitting.
The EMR data more closely resemble
a power-law with $\sigma \sim R^{-1.2}$, only slightly steeper than the whole population.
Thus in this galaxy, both GC components appear to follow rather shallow distributions, though
at large radii the EMP component dominates strongly in total numbers.
For the halo light profile, \citet{donzelli_etal2011} fit a two-component model with an
inner S\'ersic and outer exponential profile; the exponential part contains almost 80\% of
the total luminosity.  Just as for UGC 9799 and NGC 6166, there is a close match between
the EMR cluster distribution and the halo light.  

\smallskip \noindent \emph{NGC 4889:} In this case we use data in the range $F814W < 26.5$,
$1.3 < (g-I) < 2.2$.  Before carrying out radial fits to a S\'ersic function,
we attempted to assess the contribution to the GC population in the field from
the companion galaxy NGC 4886, which is a moderately large elliptical
around which a noticeable GC system is seen 
(Fig.~\ref{fig:ngc4889_xy}).  A numerical approach similar to that described in
\citet{wehner_etal2008} for the Hydra BCG NGC 3311 and its companion NGC 3309
was used here:  the field is divided up into a grid of small
$10'' \times 10''$ squares, and the observed number
of GCs within each square is assumed to be the sum of the contributions from both
galaxies combined. With the assumption that their GC systems follow Hubble-type or S\'ersic profiles,
a $\chi^2-$minimization can then be used to solve for the profile parameters
(see Wehner et al.~for details).  NGC 4886 was found to contribute negligibly to the
totals beyond a $15''$ circle centered on it, so we simply exclude that region of the
image and fit a single S\'ersic profile centered on NGC 4889 alone using the remaining
area.  

The halo light profile in $\mu_V$ as given in \citet{pahre1999} is shown as the
dotted line in Fig.~\ref{fig:ngc4889_radprof}; 
its slope matches the outer part of the EMR clusters well, though all
parts of the GC distribution are very much shallower than the halo light for
$R < 30''$ (a radius which is very nearly equal to $R_{eff}$ for the $\mu_V$ profile).
Neither metallicity component follows a simple power law well, but in rough terms
the outer regions can be described as $\sigma(EMP) \sim R^{-0.7}$ and
$\sigma(EMR) \sim R^{-1.8}$, a difference at least as large as we found for
NGC 6166 and the other BCGs.

We note that in all cases above, the halo-light radial profiles shown in the figures have been corrected for
their ellipticity (see below) as $r_{circ} = a \sqrt{1-\epsilon}$, as described in Paper II.

\begin{table*}[t]
\begin{center}
\caption{\sc Radial Profiles}
\label{tab:radial}
\begin{tabular}{lrcclccl}
\tableline\tableline\\
\multicolumn{1}{l}{Galaxy} &
\multicolumn{1}{r}{$R_e$(all)} &
\multicolumn{1}{c}{$R_e$(EMP)} &
\multicolumn{1}{c}{$R_e$(EMR)} &
\multicolumn{1}{l}{$n$(all)} &
\multicolumn{1}{c}{$n$(EMP)} &
\multicolumn{1}{c}{$n$(EMR)} &
\multicolumn{1}{l}{Note} 
\\ & (kpc) & (kpc) & (kpc) \\
\\[2mm] \tableline\\
NGC 4874 & 122 & 203 & 47 & 1.5 & 1.9 & 1.2& Cho 2016 \\
NGC 4889 & 110 & 214 & 44 & 2.6 & 1.7 & 1.7 & this paper  \\
UGC 9799 & 61 & 80 & -- & 1.4 & 1.0 & ($R^{-1.4}$) & this paper \\
UGC 10143& 114 & 120 & -- & 2.0 & 1.5 & ($R^{-1.2}$) & this paper \\
\\[1mm] \tableline
\end{tabular}
\end{center}
\vspace{0.4cm}
\end{table*}

\subsection{Azimuthal Distributions}

Another way in which the MP and MR subcomponents may reveal structural differences is in the ellipticity
$\epsilon$ of their azimuthal distribution.  In NGC 6166, we found that the mean $\epsilon$ for the EMR clusters
matched the halo light, but the EMP clusters were more nearly spherically distributed.
For the BCGs in this study, it can already be seen from the contour maps described in the previous section
that this feature appears to be a more general one.  As in Paper II the method of moments
\citep{mclaughlin_etal1994} is used to quantify the mean ellipticities $\epsilon$ of the subsystems.

For UGC 9799, clusters in the radial range $10'' < R < 100''$ were used.
For the EMP component, $\epsilon = 0.13 \pm 0.08$ and $\theta = 25^o \pm 60^o$ E of N for the position
angle of the major axis; while for the EMR component $\epsilon = 0.39 \pm 0.08, \theta =  38^o \pm 5^o$.
For all GCs combined, $\epsilon = 0.37 \pm 16, \theta = 41^o \pm 4^o$.  For the halo light, 
\citet{donzelli_etal2011} give $\epsilon = 0.35, \theta = 35.5^o$ for the outer region (the more relevant
comparison here).
Clearly the EMR clusters fairly accurately follow the halo light in both azimuthal and radial terms, while
the EMP clusters follow a distribution that is scarcely distinguishable from spherical. 

For UGC 10143, the results for GCs in the range $10'' < R < 70''$ are 
$\epsilon = 0.19 \pm 0.15$ and $\theta = 21^o \pm 62^o$ for the EMP sample;
$\epsilon = 0.51 \pm 0.23, \theta =  11^o \pm 11^o$ for the EMR sample; and lastly
for all GCs combined, $\epsilon = 0.40 \pm 0.09, \theta =  22^o \pm 10^o$.  
The ellipticity of the halo light increases rather markedly with $R$ \citep{donzelli_etal2011}, and
to trace this out we carried out our own isophotal mapping using \emph{stsdas/ellipse} on the $F475W$ image.
Over $10''-70'', \epsilon_{F475W}$ increases from 0.27 to 0.47, but the position angle stays nearly
constant at $\theta \simeq 14^o$.  Again, the EMR component matches the halo light 
within the uncertainties of the solution.  An interesting additional feature of the EMP clusters is that
their distribution appears to be somewhat asymmetric, with more of them spread to the upper right
(northeast) in Fig.~\ref{fig:ugc10143_xy2}; without having wider-field data to draw on, it is difficult
to speculate what the cause might be.

Comparable studies for other galaxies of ther azimuthal distributions for the MP and MR populations separately with
comparable sample sizes are rare, but these other studies confirm a consistent pattern for the MP population
to be distributed roughly spherically while the MR population closely follows the halo light; see
\citet{geisler_etal1996,lee_etal2008,bassino2006,harris2009b,escudero_etal2015} among others.

\subsection{Total Populations and Specific Frequencies}

In UGC 9799, the total population of GCs out to $R = 150''$ (the limit of the ACS field of view,
as well as the limit of the halo-light photometry) is $n = 9650 \pm 190$ 
brighter than $I = 27.0$ from integration of the radial profile.  This limit is 0.12 mag fainter than
the GCLF peak (Paper I) and thus should include 54\% of the total over all luminosities
assuming the GCLF is Gaussian in number per unit magnitude, and using the parameters from Paper I.
We therefore obtain $N_{tot} = 18000 \pm 400$ for $R < 150''$ (110 kpc).
Integrating the light profile \citep{seigar_etal2007} to the same radius gives a total integrated magnitude
$R_{tot} = 12.55$, which corresponds to $M_V^T \simeq -22.85$ assuming a mean color $(V-R) = 0.6$ for
giant early-type galaxies.  The specific frequency of the GC system is then
$S_N = N_{tot} \cdot 10^{0.4(M_V^T + 15)} = 13.0 \pm 0.3$.
If we choose to integrate the entire GC system profile all the way to the WFC3 field at $R = 300$ kpc,
the result would be $N_{tot} \simeq 22000$ clusters, though this estimate is much more uncertain.

For UGC 10143, if again we restrict the calculation conservatively to the ACS field of view
($R < 155''$),
the total number of GCs with $I < 27.0$ is $n = 6450 \pm 100$, which translates to
$N_{tot} = 12500 \pm 200$ over all magnitudes.  
Integration of the halo light \citep{donzelli_etal2011} to the same
radius gives $R_{tot} = 12.43$, $M_V^T \simeq -22.96$.  Finally then, $S_N = 8.2 \pm 0.1$
to that radius.  The more risky extrapolation of the GC profile out to the WFC3 field limit
at $R = 370''$ (270 kpc) would give $N_{tot} \simeq 20000$.
The UGC 9799 GC system is therefore relatively richer than for UGC 10143, but the specific frequencies of
both are in the range observed for other BCGs \citep[e.g.][]{harris_etal2013}.

For NGC 4889 the profile integration gives $n = 8080 \pm 120$ clusters brighter than 
$F814W = 26.5$ ($M_I = -8.53$), a limiting magnitude which is very near the GCLF peak
point.  We therefore double that number to $N_{tot} \simeq 16000 \pm 250$, which when
combined with $M_V^T = -23.65$ gives a specific frequency of $S_N = 5.5 \pm 0.1$.
This galaxy is very luminous, but relative to its size it does not have an exceptionally
populous cluster system in the more typical BCG range.

The story is somewhat different and more complex 
for the central Coma cD NGC 4874:  \citet{peng_etal2011} estimate
$N_{tot} = 23000 \pm 700$ GCs \emph{not} including the more extended IGC population.
If we use $M_V = -23.46$ as did \citet{harris_etal09}, we obtain $S_N = 9.5 \pm 0.3$.
If instead $N_{tot}$ is normalized to the entire luminosity profile out to a much
larger radius of 520 kpc then $S_N$ decreases markedly to $3.7 \pm 0.1$
\citep{peng_etal2011}; but since the IGC clusters outnumber the ``intrinsic'' GCs
more definitely associated with the galaxy itself, this lower specific frequency 
is perhaps more of a statement about $S_N(IGC)$.
In this respect it is worth noting that \citet{durrell_etal2014} find $S_N = 2.8 \pm 0.7$
for the Virgo cluster GCs in their entirety, of which a large fraction are IGCs and
less than a quarter are from M87 itself.

\section{Discussion}

For extremely broad and nearly featureless MDFs such we find in NGC 6166 (Paper II), UGC 9799, and UGC 10143, 
the imposition of a bimodal Gaussian numerical model, or any simplistic multimodal fitting
process, begins to look increasingly arbitrary.  Transformation of their CDFs back into 
the underlying metallicity distribution by way of
Eqn.~(1) yields a similarly broad unimodal MDF.  We stress that this is a 
different issue than 
the one raised by \citet{yoon_etal2006}, who proposed that an \emph{intrinsically
unimodal MDF} could be translated into the \emph{bimodal CDF} seen in many galaxies because of the
nonlinearity of the transformation (in their case, specifically the $(g-z)$ index, which as noted
above is among the most nonlinear of the indices in common use).  Here, we
are essentially discussing the reverse:  a CDF that is already smooth and unimodal 
cannot have come from an intrinsically bimodal MDF,
because $(g-I) \rightarrow$ [Fe/H] is (mildly) nonlinear in the sense that 
it will produce a slightly more compressed MDF rather than one that is more spread out.
The same argument would apply to any of the other color indices used in other papers
such as $(g-z), (g-i), (V-I), (B-I), (C-T_1)$, and others.

The MDF we observe today is the visible
outcome of a rapid sequence of individual GC-forming events that took place within many
halos along a hierarchical merger tree.  For a galaxy at the BCG scale, thousands of such
halos take part in this sequence at high redshift, each bringing in its own partially enriched gas.  
For a complex enough chain and a large enough
number of halos, the end product might be expected to approach a continuous MDF,
stretching from the most metal-poor GCs at an epoch where significant enrichment had
not yet occurred, to the last major rounds of GC formation at roughly Solar metallicity.
In these supergiant cases, it is evident that the merging halos over their full range
of masses contained enough gas to form large numbers of GCs at every metallicity from
[Fe/H] $\sim -1.5$ up to above Solar abundance, filling in every part of the MDF.
At later times, individual accretions of small satellites continue that will add mostly
to the metal-poor GC population in the outer halo.

Rather than attempting to reverse-engineer the formation events from properties of the
color distributions and radial distributions, ultimately 
it would be preferable to move forward from a physically based formation model
to generate true model MDFs and GC spatial distributions which can then be compared with the observed cases.  
Early steps in this direction include 
\citet{kravtsov_gnedin2005,griffen_etal2010,muratov_gnedin2010,tonini2013,li_gnedin2014,li_etal2016}, 
though each of these uses particular simplifying prescriptions for the formation 
of GCs within the halos in the merger tree.  Nevertheless, for galaxies at the highest masses
the outcome MDF in the models is broader and seen to approach a continuous distribution \citep{li_gnedin2014}.

An interesting result emerging from the BCGs, however, is that there are still notable differences
in their MDFs even at these very highest galaxy masses.  For M87 and NGC 4889 the MDFs
still show a clearly bimodal form, while others (NGC 4874, NGC 6166) have MP and MR modes that
have begun to merge, and in others (UGC 9799, UGC 10143) the MDF from the raw observations 
is smooth and unimodal, and the GMM-fitted $D-$values fall below the threshold $D \simeq 2$.
Nevertheless, in strict numerical terms a bimodal-Gaussian
deconstruction matches all these cases extremely well.  We emphasize, however
(see again Section 2), that this model for the MDF is primarily a convenient description of its
first-order features, emerging from what is intrinsically a continuous process of cluster formation.
The overall appearance of the MDF is governed primarily by the internal dispersions $\sigma_1,\sigma_2$
of the MP and MR modes and thus the amount of overlap between the modes.
By contrast, a near-uniform result we find is that the MP and MR mode centers 
$\mu_1, \mu_2$ in all the BCGs remain
separated by very similar amounts $\Delta$[Fe/H] $= (0.81 \pm 0.04)$ dex, even though their dispersions
may differ quite a lot. 

Perhaps the most interesting feature is connected with the MP and MR relative numbers ($p_1, p_2$).  
These supergiants contain roughly equal numbers of MP and MR clusters, but there are
still striking differences such as in UGC 10143, where there are relatively few MR clusters
at all radii, even in much of the inner region where the numbers of accreted MP clusters
should have been small.  This is, perhaps, an indication that the numbers of 
major gas-rich mergers from the big, metal-enriched halos 
that would have produced the metal-richer GCs may differ strongly between different
BCGs \citep[see, e.g.,][]{burke_collins2013,lidman_etal2013}.  Other possibilities are 
that by the time such mergers occurred the gas was already either largely converted to
stars or heated to the point where less
metal-rich GC formation could occur.  

Our material strongly supports the identification of the metal-richer GCs with the halo light
of their parent galaxies:  their spatial distributions are similar in both radial and azimuthal
distributions, and their overall spatial structure is consistently smoother and more regular
than the MP clusters, like the galaxy light.  Subdividing the sample into the EMP and EMR subgroups
helps to emphasize their distinctive spatial distributions.  This material adds to
to the similar evidence in other large galaxies (see Paper II), pointing to the conclusion
that the MR clusters formed along with the main stellar population of the galaxy.  
By contrast, the EMP clusters
consistently follow a distribution that in power-law terms $\sigma_{cl} \sim R^{-\alpha}$
is shallower by $0.8-1.0$ dex than the EMR clusters, very different from the structure
of the halo light and fairly close to an isothermal 
form similar to the dark-matter halos.  

The relation between gas metallicity and host galaxy mass \citep[see][]{muratov_gnedin2010,li_gnedin2014}
suggests strongly that the MP GCs formed within small, very metal-poor dwarfs. 
These GCs could therefore have accumulated either extremely early in the hierarchical chain, before
the major body of the galaxy had fully assembled;
or from accreted dwarf satellites that may come in at any later time.  By contrast, the MR
GCs should form within much bigger halos with more enriched gas.  The importance
of accretion in the buildup of the MDF that is observed today was first pointed out by
\citet{cote_etal1998,cote_etal2000} and has frequently been discussed in the later literature
\citep[for comprehensive recent discussions of the relative importance of accretions, see][]{kartha_etal2016,ferrarese_etal2016}.  
BCGs are in highly privileged locations at the dynamical centers of rich
galaxy clusters, and thus may gain the most from satellite accretions.  
The importance of their growth by dissipationless (dry) merging especially for redshifts $z < 1$
has been emphasized in numerous recent discussions            
\citep[e.g.][among many others]{liu_etal2013,laporte_etal2013,laporte_etal2015,oliva-altamirano_etal2015,oogi_etal2016}.
The individual satellites
may contribute both their own GCs and their stripped nuclei, which are structurally similar
to luminous GCs.  For example, \citet{ferrarese_etal2016} calculate that almost 40\% of the GCs in the 
core region of the Virgo cluster around M87 may come from former satellites.
A potentially related pattern that may be emerging from our BCG study is that the relative dominance
of the MP clusters increases markedly beyond $R \gtrsim 4 R_{eff}$.  This rough transition
may mark the characteristic radius beyond which late accretions of metal-poor 
satellites dominate the GC population.

\section{Summary}

Continuing our series of studies of the extraordinarily rich globular cluster 
systems around Brightest Cluster Galaxies
(BCGs) with the HST ACS and WFC3 cameras, we present new comprehensive photometric analyses of the 
GC systems around UGC 9799 and UGC 10143, along with comparison data for the 
Coma supergiants NGC 4874 and NGC 4889.
Our principal findings are these:
\begin{enumerate}
\item{} The GC systems in all these galaxies are as expected extremely populous, yielding
	total populations of anywhere from 12000 to 23000 clusters within 
	galactocentric radii $\lesssim 120$ kpc.  Extrapolation to larger radii
	might almost double those totals, though the presence of as-yet unknown numbers
	of intragalactic GCs will come into play there.
\item{} The color distribution (CDF) of the GCs has been measured in the $(F475W-F814W) \simeq
	(g-I)$ color index.  In all the BCGs the CDF is broad, nominally unimodal, and
	skewed (asymmetric).  Nevertheless,
	a simple bimodal-Gaussian deconstruction continues to match the CDF
	very well, as it has for most smaller galaxies.  The primary difference between
	these BCGs and smaller galaxies is that the intrinsic dispersions of the MP and MR
	modes become significantly higher, forcing the two modes to overlap heavily.
\item{} The broad, near-continuous form of the CDF implies that the intrinsic metallicity
	distribution function (MDF) must also be broad and unimodal, because the transformation
	from color to metallicity is only slightly nonlinear and acts in the direction
	of making the MDF slightly more compressed than the CDF.
\item{} Of the four galaxies discussed here, the Coma giants NGC 4874 and NGC 4889 show a
	mass-metallicity relation (MMR) along the MP sequence, in the conventional sense
	that the blue sequence becomes systematically redder at higher luminosity.
	The heavy-element abundance scaling with GC mass is $Z \sim M^{0.25}$,
	which was also the case for NGC 6166 (Paper II).
	For UGC 10143, no detectable trend shows up along either
	the red or blue sequences.  For UGC 9799, the results are uncertain, with 
	either a zero or positive slope along the blue sequence not ruled out.
	These results do not appear to support a simple self-enrichment model during
	GC formation, and instead may point to the need for some form of pre-enrichment
	in the most massive GCs driven by environmental differences at the time of formation.
\item{} The relative numbers of MP and MR clusters within these supergiants
	differ significantly in detail, with at least one case
	(UGC 10143) where the MP clusters dominate at all radii.  This result suggests that the relative
	importance of gas-rich mergers with big, metal-enriched halos (which built the MR clusters)
	could have differed between BCGs.
\item{} The ratio N(MP)/N(tot) increases with radius in all these galaxies, particularly
	past $R \gtrsim 4 R_{eff}$.  We suggest tentatively that this transition radius
	may mark the region outside which the MP clusters that came from
	late, discrete accretions of dwarf satellite galaxies are most important.
\item{} In all these BCGs, as in other large galaxies, the extremely metal-poor GCs follow a roughly spherical
	spatial distribution not far from the form $\sigma \sim R^{-1}$ that would
	characterize the dark-matter halo.  By contrast, the most metal-rich GCs follow
	a more concentrated, smooth, and regular distribution that matches the halo
	light of the galaxy. The MR clusters are most likely to have formed along with the
	main stellar body of the galaxy.
\end{enumerate}

\section*{Acknowledgements}

Based on observations made with the NASA/ESA Hubble Space Telescope, obtained 
at the Space Telescope Science Institute, 
which is operated by the Association of Universities for 
Research in Astronomy, Inc., under NASA contract NAS 5-26555. 
WEH and GME acknowledge financial support 
from NSERC (Natural Sciences and Engineering Research Council of Canada).
JB acknowledges financial support from program HST-AR-13908.001-A 
provided by NASA through a grant from the Space Telescope Science Institute.
OYG was supported in part by NASA through grant NNX12AG44G, and by NSF through grant 1412144.
DG gratefully acknowledges support from the Chilean BASAL Centro de
Excelencia en Astrof\'isica y Tecnolog\'ias Afines (CATA) grant PFB-06/2007.  

{\it Facilities:} \facility{HST (ACS, WFC3)}

\makeatletter\@chicagotrue\makeatother

\bibliographystyle{apj}
%\bibliography{gc}

%\label{lastpage}

\end{document}